\documentclass[twocolumn,citeautoscript,pra,superscriptaddress,amsmath,amssymb,8pt]{revtex4-1}

\usepackage{graphicx}
\usepackage{upgreek}
\usepackage{float}
\usepackage[T1]{fontenc}

\usepackage{hyperref}
\hypersetup{
	colorlinks=true,
	linkcolor=blue,
	filecolor=blue,      
	urlcolor=blue,
	citecolor=blue
}

\usepackage{tcolorbox}
\tcbuselibrary{breakable}

\usepackage{tabularx}
\newcolumntype{C}[1]{>{\centering\arraybackslash}p{#1}}

\newcommand{\vect}[1]{\ensuremath{\textbf{#1}}}

\newcommand{\abs}[1]{\ensuremath{\left| #1 \right|}}
\newcommand{\magunit}[0]{\ensuremath{\mu_B\text{nm}^{-2}}}
\newcommand{\FJ}[0]{\ensuremath{\mathcal{J}}}

\newcommand{\unv}[0]{\ensuremath{\mathbf{u}~}}
\newcommand{\angles}[2]{\ensuremath{\ensuremath{(\theta, \phi) = (#1^\circ, #2^\circ)}}}

\newcommand{\BNV}[0]{\ensuremath{B_{\theta,\phi}}}
\newcommand{\FBNV}[0]{\ensuremath{\mathcal{B}_{\theta, \phi}~}}

\begin{document}

\title{Improved current density and magnetisation reconstruction through vector magnetic field measurements}

\author{D. A. Broadway}
\email{davidaaron.broadway@unibas.ch}
\affiliation{School of Physics, University of Melbourne, Parkville, VIC 3010, Australia}
\affiliation{Centre for Quantum Computation and Communication Technology, School of Physics, University of Melbourne, Parkville, VIC 3010, Australia}
\affiliation{Department of Physics, University of Basel, Klingelbergstrasse 82, Basel CH-4056, Switzerland.}

\author{S. E. Lillie}
\affiliation{School of Physics, University of Melbourne, Parkville, VIC 3010, Australia}
\affiliation{Centre for Quantum Computation and Communication Technology, School of Physics, University of Melbourne, Parkville, VIC 3010, Australia}

\author{Sam C. Scholten}
\affiliation{School of Physics, University of Melbourne, Parkville, VIC 3010, Australia}
\affiliation{Centre for Quantum Computation and Communication Technology, School of Physics, University of Melbourne, Parkville, VIC 3010, Australia}

\author{D. Rohner}
\affiliation{Department of Physics, University of Basel, Klingelbergstrasse 82, Basel CH-4056, Switzerland.}

\author{N. Dontschuk}
\affiliation{School of Physics, University of Melbourne, Parkville, VIC 3010, Australia}

\author{P. Maletinsky}
\affiliation{Department of Physics, University of Basel, Klingelbergstrasse 82, Basel CH-4056, Switzerland.}

\author{J.-P. Tetienne}
\email{jtetienne@unimelb.edu.au}
\affiliation{School of Physics, University of Melbourne, Parkville, VIC 3010, Australia}
\affiliation{Centre for Quantum Computation and Communication Technology, School of Physics, University of Melbourne, Parkville, VIC 3010, Australia}

\author{L. C. L. Hollenberg}
\email{lloydch@unimelb.edu.au}
\affiliation{School of Physics, University of Melbourne, Parkville, VIC 3010, Australia}
\affiliation{Centre for Quantum Computation and Communication Technology, School of Physics, University of Melbourne, Parkville, VIC 3010, Australia}

\begin{abstract}
	Stray magnetic fields contain significant information about the electronic and magnetic properties of condensed matter systems. For two-dimensional (2D) systems, stray field measurements can even allow full determination of the source quantity. For instance, a 2D map of the stray magnetic field can be uniquely transformed into the 2D current density that gave rise to the field and, under some conditions, into the equivalent 2D magnetisation. However, implementing these transformations typically requires truncation of the initial data and involves singularities that may introduce errors, artefacts, and amplify noise. Here we investigate the possibility of mitigating these issues through vector measurements. For each scenario (current reconstruction and magnetisation reconstruction) the different possible reconstruction pathways are analysed and their performances compared. In particular, we find that the simultaneous measurement of both in-plane components ($B_x$ and $B_y$) enables near-ideal reconstruction of the current density, without singularity or truncation artefacts, which constitutes a significant improvement over reconstruction based on a single component (e.g. $B_z$). On the other hand, for magnetisation reconstruction, a single measurement of the out-of-plane field ($B_z$) is generally the best choice, regardless of the magnetisation direction. We verify these findings experimentally using nitrogen-vacancy magnetometry in the case of a 2D current density and a 2D magnet with perpendicular magnetisation.
\end{abstract}

\maketitle

\section{Introduction}

Condensed matter systems are often accompanied by stray magnetics fields that are the result of uncompensated magnetic moments or the movement of charges in the material. Local magnetic field measurements can therefore be used to investigate the underlying physical phenomena. Stray magnetic fields can be measured using a suitable magnetic probe such as superconducting quantum interference device (SQUID)~\cite{Shi2019,Vasyukov2013,Kalisky2013}, Hall probe~\cite{Nowack2013,Dinner2007,Dinner2007a}, and nitrogen-vacancy (NV) centres in diamond~\cite{Doherty2013}, arranged either in a dense array or scanning configuration to form a two-dimensional (2D) magnetic field map~\cite{Rondin2014, Rondin2012,  Maletinsky2012, Steinert2010, Pham2011, Tetienne2017a}. Using these techniques various physical phenomena have been investigated such as  quantum Hall effects~\cite{Uri2019, Nowack2013}, spin wave modes~\cite{Du2017}, magnetism at oxide interfaces~\cite{Anahory2016}, 2D magnetic materials~\cite{Gibertini2019, Gong2019, Thiel2019, Broadway2020, Wornle2019, Kim2019}, non-collinear magnetism~\cite{Gross2017a}, and domain wall physics~\cite{Tetienne2014,Tetienne2015}. Although less explored, it is also a promising avenue to study transport phenomena such as viscous electron flow~\cite{Bandurin2016}, electron guiding and lensing~\cite{Cheianov2007,Chen2016b}, spintronics~\cite{Jungwirth2016}, and topological currents~\cite{Gorbachev2014}.

In some cases, it is possible in principle to transform a 2D map of the stray magnetic field into the source quantity~\cite{Roth1989, Meltzer2017, Knauss2001, Clement2019, Lima2009, Casola2018, Tan1996, Dreyer2007, Thiel2019, Broadway2020}. Specifically, a 2D measurement of the stray magnetic field (by convention, in the $xy$-plane parallel to the sample) can be transformed to obtain a map of 2D current density and, under some conditions, of 2D magnetisation. Typically, these 2D measurements are performed such that only a single projection of the stray magnetic field is measured, which is to decrease the total measurement time (NV~\cite{Tetienne2017a}) or due to the difficulty in designing a 3-axis probe (Hall probe, SQUID~\cite{Anahory2014}). However, the transformation from magnetic field to source quantity contains singularities (undetermined spectral components) that vary depending on the magnetic field direction in question. For example, the transformation from perpendicular magnetic field ($B_z$) to a 2D out-of-plane magnetisation ($M_z$) contains a single point singularity (the DC Fourier frequency pixel is divided by zero) while the transformation from a transverse magnetic field component ($B_x$ or $B_y$) contains many singularities. Likewise, truncation artefacts arising from the finite lateral size of the measurements (in the $xy$ plane) are more or less pronounced depending on the measurement direction. Recently, it has been possible to measure the full vector magnetic field directly using ensembles of NV centres in diamond~\cite{Steinert2010,Pham2011,Maertz2010}, which opens new opportunities to mitigate these effects in a systematic manner. Here we investigate how this increased information can be used to choose a transformation pathway (e.g. from $\mathbf{B}$ to $\mathbf{M}$) that minimises singularity and truncation induced artefacts.   

Several techniques exist to transform the magnetic field into the desired source quantity such as Bayesian inference~\cite{Clement2019}, regularization~\cite{Feldmann2004,Meltzer2017}, and Fourier-space reconstruction~\cite{Roth1989, Tetienne2019, Lima2009, Casola2018}. In this work, we focus on the simplest method of Fourier-space source quantity reconstruction. We compare the use of different magnetic field projections to reconstruct current density, in-plane magnetisation, and out-of-plane magnetisation. Namely, we analyse the use of Cartesian projections $B_x$, $B_y$, $B_z$, and the projection along an arbitrary direction \BNV, as well as combinations of these magnetic field components. First, in section~\ref{Sec: curr}, these different magnetic field components are explored for current density reconstruction for non-trivial current geometries and for regimes where there is a high degree of magnetic field truncation.  In section~\ref{Sec: exp curr}, these reconstruction methods are applied to experimental data from a metallic wire fabricated onto an NV-diamond magnetic imager revealing a clear advantage in using $B_x$ and $B_y$ together. Then  in section~\ref{Sec: mag}, we discuss the transformation from magnetic fields to magnetisation and apply these transformations to out-of-plane magnetisation in section~\ref{Sec: out-of-plane}. Here we show that transformation involving transverse magnetic fields have a distinct disadvantage to $B_z$ and in section~\ref{Sec: exp out-of-plane} the results are confirmed by implementing the different reconstruction pathways on experimental data of magnetic flakes of vanadium triiodide (VI$_3$). Finally, in section~\ref{Sec: in plane}, we investigate in-plane magnetisation returning the same conclusion as for out-of-plane, that $B_z$ is the preferred measurement orientation. 

\section{Reconstruction of current density}\label{Sec: curr}

We first examine the case of current density reconstruction. Here we will show that it is possible to use the in-plane field components together to get a singularity free reconstruction, while using the out-of-plane field leads to a singularity in the transformation. Additionally, we show that there are further artefacts that are introduced due to the longer range of decay of perpendicular magnetic field as compared to the in-plane components for the systems of interest. We demonstrate that these artefacts can change the apparent distribution of the current density in the source material and introduce edge artefacts. Finally, we demonstrate this variation on experimental data where a clear difference in reconstruction from perpendicular and transverse fields is observed. 

\subsection{Theory}\label{Sec: B to curr transformation}

\begin{figure*}
	\includegraphics[width=\textwidth]{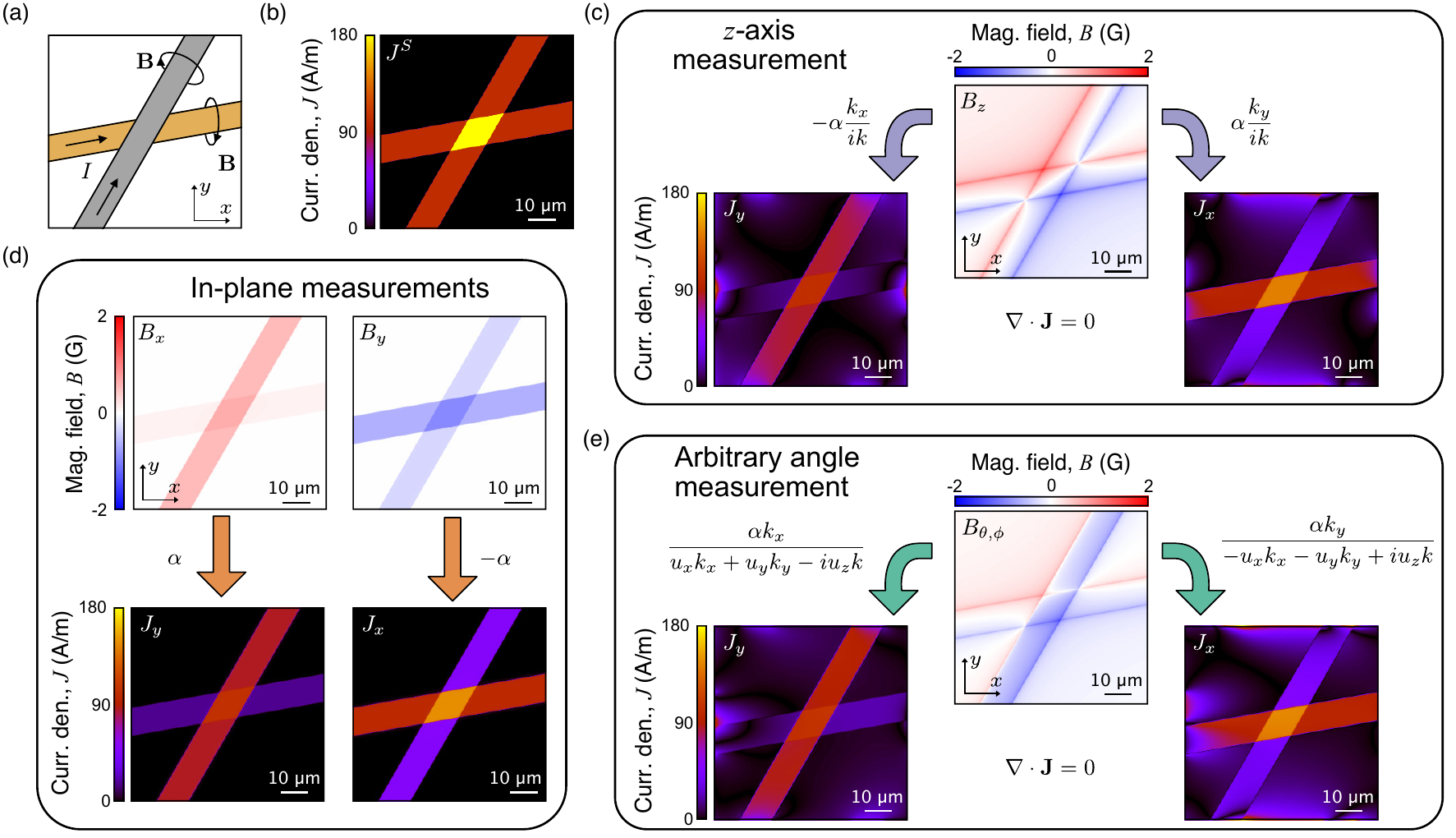}
	\caption{ \textbf{Reconstruction of current density.} 
		(a) Diagram of two independent current carrying wires (same current $I$ in each) in non-trivial geometry. 
		(b) Simulated current density corresponding to the situation shown in (a) for a current $I=2$~mA in each wire.
		(c) Calculated $B_z$ magnetic field (top panel) then used to reconstruct both $J_x$ and $J_y$ current densities (bottom panels) based on Eq.~\ref{Eq: Bz to J} which assumes that $\nabla \cdot \mathbf{J} = 0$, where the magnitude ($|J_{x,y}|$) is shown for clarity. 
		(d) Calculated $B_x$ and $B_y$ maps (top panels) used to reconstruct the current density (bottom panels) based on Eq.~\ref{Eq: Bxy to J}.
		(e) Calculated \BNV~map (top panel) used to reconstruct the current density (bottom panels) based on Eq.~\ref{Eq: Bnv to J}, where \BNV~magnetic field was simulated using \angles{54.7}{45}.
	}
	\label{Fig: current propagation}
\end{figure*}

	The reconstruction of current density from the measured magnetic field can be performed provided the magnetic field is measured in an $xy$-plane parallel to the confinement plane of the 2D current density (defined as the $z=0$ plane), with a known height $z'>0$. Because the current is confined in 2D, the current density has only two vector components, ${\bf J}=(J_x,J_y,0)$. The magnetic field ${\bf B}$ is related to ${\bf J}$ via the Biot-Savart equation,
	\begin{equation}
	\mathbf{B}( \mathbf{r}) = \frac{\mu_0}{4\pi} \int d^3 \mathbf{r}' \frac{ \mathbf{J} ( \mathbf{r}' ) \times (\mathbf{r} - \mathbf{r}')}{\abs{\mathbf{r} - \mathbf{r}'}^3}, 
	\end{equation} 
	where $\mu_0$ is the vacuum permeability and the integration is over all space. This relationship is simplified in Fourier-space where the Cartesian components are related by
	\begin{equation} \label{Eq: curr transform simp}
	\begin{bmatrix} \mathcal{B}_x \\  \mathcal{B}_y \\ \mathcal{B}_z \end{bmatrix} 
	=	\frac{1}{\alpha}
	\begin{bmatrix} 0& 1  \\   - 1 & 0 \\ i k_y / k & - i k_x / k \end{bmatrix} 
	\begin{bmatrix} \mathcal{J}_x \\  \mathcal{J}_y  \end{bmatrix},
	\end{equation}
	where $\mathcal{J}(k_x, k_y)$ is the Fourier-space current density, $\mathcal{B}(k_x, k_y, z')$ is the Fourier-space magnetic field where $k_x$ and $k_y$ are the Fourier-space vector variables with $k = \sqrt{k_x^2 + k_y^2}$. The term $\alpha = 2e^{k z'} /\mu_0$ contains an exponential factor that includes the stand-off from the source $z'$, which acts to reverse propagate the magnetic field to the source plane~\cite{Lima2009}.  When all three field components are known, Eq.~\ref{Eq: curr transform simp} is an overconstrained problem, and so there are several ways to deduce $\mathcal{J}$. An obvious pathway is to use the in-plane field components only ($B_x$ and $B_y$), which are trivially related to $J_x$ and $J_y$ giving the inverted transformation 
	\begin{equation} \label{Eq: Bxy to J}
	(B_x,~ B_y) \rightarrow
	\left\lbrace
		\begin{aligned}
			\mathcal{J}_x &=  - \alpha \mathcal{B}_y, \\
			\mathcal{J}_y &=  \alpha \mathcal{B}_x. 
		\end{aligned}
		\right.
	\end{equation}
 The inversion expressed by Eq.~\ref{Eq: Bxy to J} is complete and contains no singularities, that is, it is defined in the entire $k$-space. An alternative reconstruction pathway is to use $B_z$ only along with the additional condition of $\nabla \cdot \mathbf{J} = 0$ (continuity of current) which gives
	\begin{equation} \label{Eq: Bz to J}
		B_z \rightarrow
		\left\lbrace
		\begin{aligned}
			\mathcal{J}_x  &=  \frac{\alpha k_y}{ik} \mathcal{B}_z, \\
			\mathcal{J}_y  &=  - \frac{\alpha k_x}{ik} \mathcal{B}_z.
		\end{aligned}
		\right.
	\end{equation}
	Although this is the most commonly employed pathway~\cite{Nowack2013, Roth1989}, one downside is that there exists a singularity at $k=0$ where $\mathcal{J}$ is undetermined. In the real-space, this corresponds to an unknown overall DC offset of the {\bf J} map. For a single field component \BNV~along an arbitrary direction  \unv  $= (u_x, u_y, u_z) = (\sin\theta\cos\phi, \sin\theta\sin\phi,\cos\theta)$, Eq.~\ref{Eq: Bz to J} can be generalised,
	\begin{equation}\label{Eq: Bnv to J}
			\BNV \rightarrow
		\left\lbrace
		\begin{aligned}
		\mathcal{J}_x &=  \frac{\alpha k_y}{-u_x k_x - u_y k_y + i u_z k} \FBNV, \\
		\mathcal{J}_y &=    \frac{ \alpha k_x}{u_x k_x + u_y k_y - i u_z k} \FBNV.
		\end{aligned}
		\right.
	\end{equation}
	where \FBNV is the Fourier transform of \BNV~and ($\theta,\phi$) are the spherical angles. Here the transformation in general has the same singularity as the transformation for $B_z$, that is, $\mathcal{J}$ is undetermined when $k=0$. In the case of a purely in-plane measurement, e.g. along $x$ i.e. \unv $= (1,0,0)$, Eq.~\ref{Eq: Bnv to J} becomes $\FJ_x = -( \alpha k_y/k_x ) B_x$, $\FJ_y = \alpha B_x$. The expression for $\FJ_y$ is the same as in Eq.~\ref{Eq: Bxy to J}, however $\FJ_x$ now has many singularities (when $k_x = 0$) stemming from the reduction in available information, seriously compromising the reconstruction. This indicates that, when a single magnetic field projection is measured, it is preferable if this projection has a significant $z$-component to minimise the impact of singularities.  
	
	Another aspect to consider is finite-size effects. Indeed, $\mathbf{B}$ is measured over a finite spatial region (in the $xy$-plane) near the sample, which leads to artefacts in the reconstructed $\mathbf{J}$ when using the above equations. These artefacts are sometimes known as truncation artefacts~\cite{Tetienne2019}. To analyse how these vary depending on which $\mathbf{B}$ component is used, we simulate a generic scenario of two current carrying wires with an angle of $\theta = 10^\circ$ and $60^\circ$ from the $x$-axis and a stand-off of $z' = 50$~nm (Fig.~\ref{Fig: current propagation}a,b). The simulation has a pixel size of 100~nm where a Gaussian smoothing was applied with a width of 300~nm to mimic the diffraction limit of optical imaging techniques~\cite{Tetienne2017a}. While the simulations are performed under these conditions, the results are also valid for scanning systems which have a higher spatial resolution.
	  
\begin{figure*}
	\includegraphics[width=\textwidth]{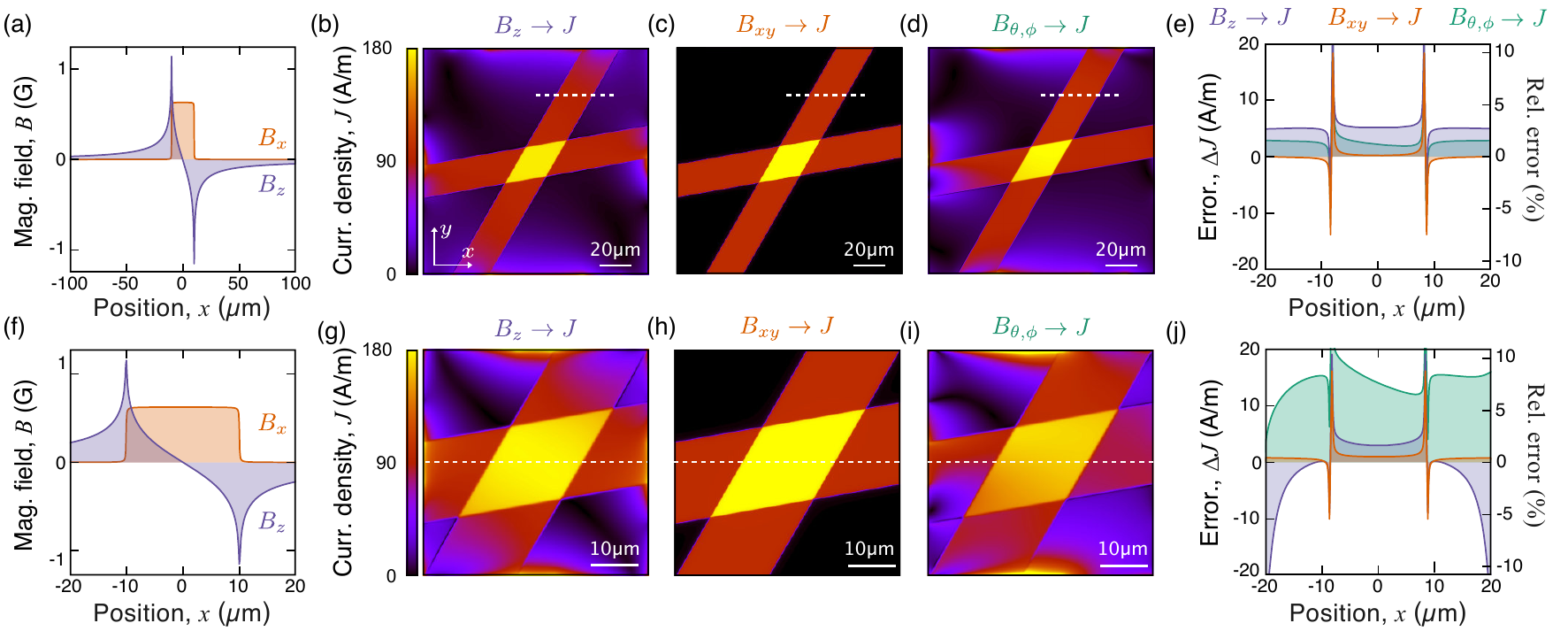}
	\caption{\textbf{Truncation errors in current density reconstruction.} 
		(a) Simulated magnetic field from a 20~$\mu$m wide wire with a stand-off of 50~nm, orientated along the $y$-direction with a small amount of spatial truncation of $B_z$. 
		(b-d) Current density norm ($J$) reconstructed from $B_{z}$ (b), $B_{xy}$ (c), and \BNV~(d) of the wire geometries in Fig.~\ref{Fig: current propagation}. 
		(e) Line cuts of the difference between the simulated current density and the reconstructed current density for different sources, taken along the dashed line in (b-d). The right axis is the relative error normalised to the maximum simulated current density. 
		(f) Magnetic field with a reduced image size.
		(g-i) Reconstructed current density for a measurement confined to the smaller imaging size for $B_z$ (g), $B_{xy}$ (h), and \BNV~(i).
		(j) Same as (e) but with the reduced image size, taken along the dashed line in (g-i). The \BNV~magnetic field was simulated using \angles{54.7}{45}.
	}
	\label{Fig: curr truncation}
\end{figure*}

The simulated $B_z$ magnetic field (Fig.~\ref{Fig: current propagation}c, top panel) is transformed in Fourier-space into current density, where initially the $k=0$ component is set to zero. Then, back in real-space, a DC offset is added such that the current density vanishes far from the wires (precisely, we impose that the average current density near the top-left corner be zero). Both the transformation to $J_x$ (Fig.~\ref{Fig: current propagation}c, bottom right panel) and $J_y$ (Fig.~\ref{Fig: current propagation}c, bottom left panel) have significant edge related reconstruction errors. The edge artefacts are related to the way that $B_z$ extends relative to the wire, that is, $B_z$ fields persist significantly beyond the finite window of the measurement. Thus any abrupt end to the magnetic image introduces errors, either along the wire itself, or in the extended field which introduces magnetic field truncation, which is discussed in the next section. In contrast, the reconstruction from the $B_x$ (Fig.~\ref{Fig: current propagation}d, left panels) and $B_y$ (Fig.~\ref{Fig: current propagation}d, right panels) magnetic fields (collectively referred to as $B_{xy}$) is free of such artefacts. However, in order to perform this reconstruction one either needs to take multiple measurements or use a probe that is sensitive to multiple magnetic field orientation at once, which comes at the cost of signal to noise. 

A single measurement with an arbitrary orientation with respect to the sample can also be used to reconstruct the current density. Here we simulate the case of \angles{54.7}{45} which is a common situation in NV magnetometry using a (100)-oriented diamond (Fig.~\ref{Fig: current propagation}e). However, in this case the measurement suffers from the same truncation errors from the $B_z$ component and the reduced information from the in-plane component. Leading to a very similar current distribution to that of the $B_z$ reconstructed current density.

The truncation affects the low-$k$ components (such that $kz' \ll 1$) which are coarse grained as a result. The transformation for $B_z$ has a $1/k$ factor, which means errors in $B_z$ (low $k$) are amplified. Meanwhile, the transformation for transverse fields only has the $e^{kz'}$ factor, which is very close to 1 for low $k$ and as such introduces very few truncation artefacts. This is also true in the case of scanning probes because the size of the image (lowest $k$) is always much larger than the stand-off.

\subsection{Truncation induced errors}\label{Sec: curr trunctation}

We now analyse these truncation errors in more detail. The truncation of the $B_z$ data depends on the image size in comparison with the size of the wire. This is particularly relevant for scanning sensor systems that typically use small image sizes limited by the long acquisition times involved. While these scanning sensor systems have exceptional resolution for probing small devices~\cite{Maletinsky2012, Pelliccione2016, Thiel2016, Chang2017}, when applied to larger structures the smaller field of view may result in significant truncation. Due to the slow decay ($1/r$) of the $B_z$ field it is rare to have no truncation and as such in most measurements there will be some truncation of the magnetic field (Fig.~\ref{Fig: curr truncation}a). 

The reconstruction of the total current density from the simulated $B_z$ map (Fig.~\ref{Fig: curr truncation}b) is plagued with edge artefacts due to the truncation of the  $B_z$ field. In contrast, the reconstruction from the $B_{xy}$ maps (Fig.~\ref{Fig: curr truncation}c) has no artefacts and doesn't require additional processing to improve the reconstruction. The reconstruction from a single arbitrary orientation \angles{54.7}{45} shows a similar error to the case with pure $B_z$, where they both experience an offset in the wire and outside (Fig.~\ref{Fig: curr truncation}d). However, the reconstruction with \BNV~also results in a non physical gradient in the wire. Line cuts of the difference between the simulated current ($J^S$) and the reconstructed current ($J^R$), $\Delta J = J^S - J^R$, across one of the wires show that even away from the edge effects the reconstruction from $B_{xy}$ more accurately produces the simulated current density than the $B_z$ and \BNV~reconstructions (Fig.~\ref{Fig: curr truncation}e). 

When the image size is reduced further (Fig.~\ref{Fig: curr truncation}f) the error in reconstruction from $B_z$ increases, particularly near the edges of the image (Fig.~\ref{Fig: curr truncation}g), while the $B_{xy}$ transformation is unchanged (Fig.~\ref{Fig: curr truncation}h) and only has issues with edges of the wire due to the finite pixel size in the measurement. The \BNV~reconstruction has an even worse response than the straight $B_z$ reconstruction, due to an asymmetry that is introduced in the reconstructed current density (Fig.~\ref{Fig: curr truncation}i). The difference from the simulated current density (Fig.~\ref{Fig: curr truncation}j) taken along the centre of the image shows that reconstruction with $B_z$ can have a deviation of up to 3\% in these conditions and \BNV~greater than 10\%, relative to the total expected current density (compared to less than 1\% in the bulk of the wire for $B_{xy}$). These deviations are significant and need to be considered when investigating small current effects like edge modes, or spin contributions. Additionally, we performed similar simulations including some noise in the $\mathbf{B}$ data, and found no significant modification of the different reconstruction pathways with the introduction of noise, i.e. the different pathways all returned a similar SNR after reconstruction. We also note that there has been extensive work on how to mitigate the effect of noise in the case of single measurement axis~\cite{Meltzer2017,Clement2019}. 

In trivial geometries (e.g. single wire in the $x$-direction) the truncation of the magnetic field can be mitigated completely through fitting the magnetic field directly rather than reconstructing the current density~\cite{Ku2019}. An alternative method involves padding of the real-space image with a linear or exponential decay to extrapolate the data outside the measurement window~\cite{Tetienne2019, Chang2017}. However, in more complicated geometries this padding is not reliable as the extrapolation introduces additional artefacts. Likewise, simulations of the magnetic tails may involve incorrect assumptions about current distributions and lead to erroneous results. A compromise is to include padding with zeros, which effectively halves the errors due to truncation of the $B_z$ component~\cite{Tetienne2019}. 

A consequence of the truncation artefacts is that they produce a non-uniform background current density where there should be no current at all, which makes it difficult to determine the DC offset in the $\mathbf{J}$ maps, and as a result may bias the estimated current density in the wires themselves. Therefore, it is typically necessary to manually choose a region for zero-point normalisation that looks the most artefact free. For instance, in Fig.~\ref{Fig: curr truncation} we chose the top-left corner of the image. Thus, compared with reconstruction from $B_{xy}$, which requires no normalisation and produces an accurate estimation of the actual current density, reconstruction from $B_z$ and \BNV~is left wanting.

\subsection{Experimental comparison}\label{Sec: exp curr}

\begin{figure*}
	\includegraphics[width=\textwidth]{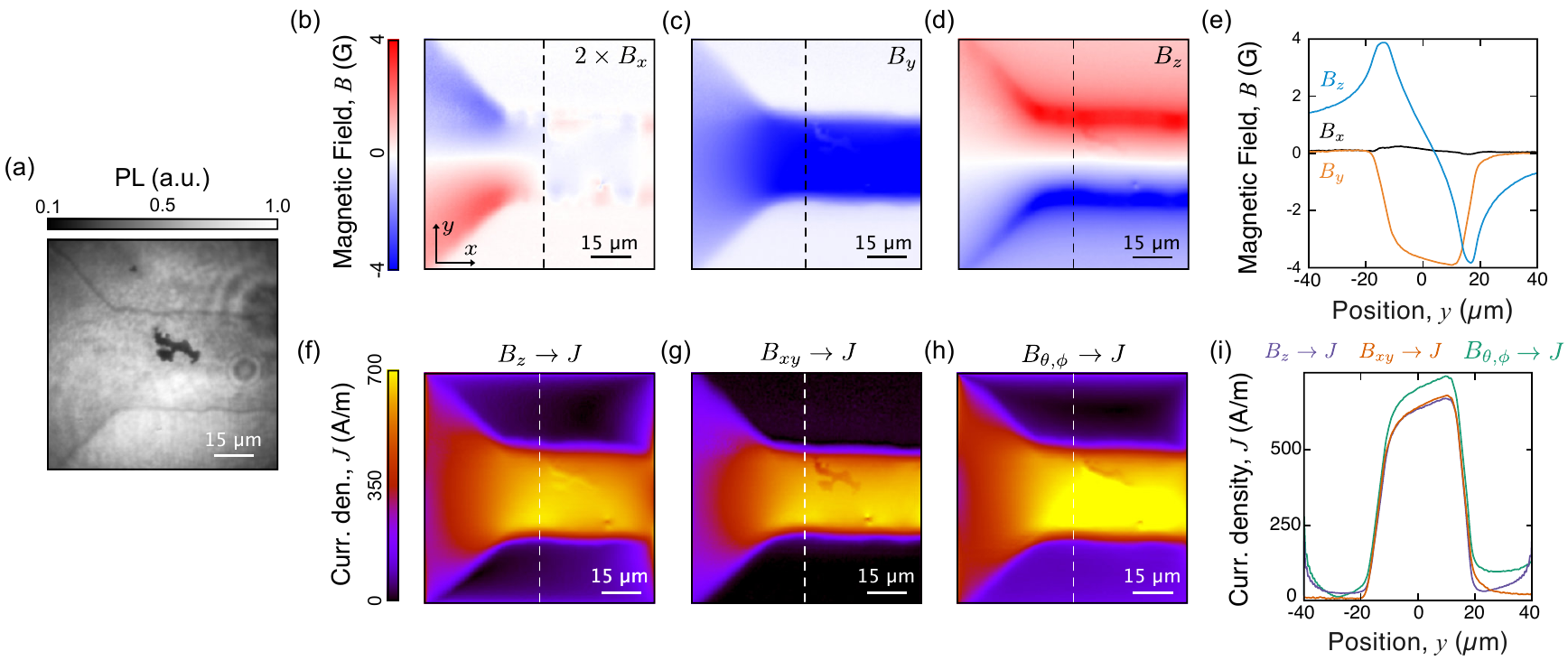}
	\caption{\textbf{Experimental example of current reconstruction.} 
		(a) NV photoluminescence image of a Nb wire fabricated directly onto the diamond surface. 
		(b-d) Magnetic field images from the Nb wire with a current of $I = 20$~mA deduced from optically detected magnetic resonance (ODMR) spectroscopy. 
		(e) Line cuts of the different magnetic fields, taken along the black dashed line in (b-d).
		(f-h) Reconstructed current density norm, $J$, from $B_{z}$ (f), $B_{xy}$ (g) and \BNV~(h) magnetic fields. Where the reconstruction with $B_{xy}$ has no padding or additional normalisation. Conversely, the reconstruction with $B_{z}$ and \BNV~is padded with zeros and has a DC offset adjusted in order to cancel $J$ near (but not exactly at) the top edge of the image (see text).
		(i) Line cuts of the reconstructed current density from different magnetic field sources, taken along the white dashed line in (f-h). 
		The \BNV~magnetic field used had an approximate angle of $(\theta, \phi) \approx (54.7^\circ, 45^\circ)$. 
	}
	\label{Fig: Exp curr}
\end{figure*}

Experimentally, we validate the simulations using a niobium (Nb) wire fabricated directly on the diamond surface (see photoluminescence image in Fig.~\ref{Fig: Exp curr}a). The Nb wire had a thickness of 200~nm and consisted of two 200~$\mu$m wide bonding pads that narrow down to a 40~$\mu$m wide channel between them~\cite{Lillie2020}. In order to produce a strong magnetic field, a current of $I = 20$~mA was used.  The measurements are taken at room temperature using an ensemble of NV spins~\cite{Tetienne2018a} with a depth distribution peaking at 120~nm below the diamond surface~\cite{Lillie2020}. The Cartesian magnetic field components were obtained through fitting the Hamiltonian to optically detected magnetic resonance (ODMR) of all four NV orientations~\cite{Tetienne2017a,Broadway2018b,Broadway2018c}, where a small bias field of $B=100$~G was applied to distinctly split all four NV orientations. The imaging was performed with a custom built widefield fluorescence microscope~\cite{Lillie2020,Simpson2016, LeSage2013, Levine}. The vector magnetic field components are shown in Fig.~\ref{Fig: Exp curr}b-d where a reference measurement is used ($I=0$) to remove the applied background field. Now we investigate the different current reconstruction pathways with this dataset. 

With an image size of $80~\mu$m $\times 80~\mu$m (only twice the width of the wire) the transverse magnetic fields (Fig.~\ref{Fig: Exp curr}b,c) are captured completely while the $B_z$ magnetic field (Fig.~\ref{Fig: Exp curr}d) is truncated significantly. This is quite explicit in the magnetic field line cuts across (Fig.~\ref{Fig: Exp curr}e) that show the $B_z$ is still above 25\% of the maximum value at the edge of the image. While the current reconstruction from transverse fields (Fig.~\ref{Fig: Exp curr}f) results in a clear current map with $J=0$ outside of the wire, the reconstruction from $B_z$ (Fig.~\ref{Fig: Exp curr}g) has a non-zero background as well as artefacts at the edge of the image.  These results are consistent with the simulations in the previous section. 

To compare to reconstruction from a single arbitrary direction \BNV~measurement, the same data set was used and the magnetic field map from a single NV orientation before conversion to $B_{xyz}$ was employed, where the \BNV~map with the least relative noise was chosen (here $ (\theta, \phi) \approx (54.7^\circ, 45^\circ)$, although the results were consistent independent of the NV orientation). The reconstruction from \BNV~produces similar artefacts to the $B_z$ case with additional gradients caused by the asymmetry in the \BNV~map, leading to a highly non-uniform background (Fig.~\ref{Fig: Exp curr}h). 

To analyse these effects more quantitatively, we look at line cuts taken through the middle of the image  (Fig.~\ref{Fig: Exp curr}i), i.e. away from the edge artefacts. In this regime, the major difference between the different pathways is the amount of current that is attributed to the background, which ultimately changes the integrated current measured depending on the normalisation protocol. The total integrated current for the reconstruction from $B_{xy}$ is $I=20(1)$~mA, which is consistent with the applied $I=20$~mA. In contrast, the integrated current measured for reconstruction from $B_z$ requires a normalisation. To deal with the current outside of the wire one can force $J=0$ near the edge of the image (zero-point normalisation) resulting in a total current density of $I = 21(1)$~mA, which is consistent with the applied value with a slight over estimation due to the edge artefacts. Unlike the reconstruction from $B_z$, the background from the \BNV~recostruction is asymmetric about the wire. As a consequence, background normalisation becomes even more difficult and even with setting a zero point in the upper background the total integrated current is $I = 24(1)$~mA, which is far from the applied current of $I=20$~mA. Assumptions about the wire can be used to negate this background in the reconstruction itself, e.g. extrapolation of the magnetic field before transformation, or a background fitting technique can be used to remove the residual current density after the transformation, but none of these techniques are perfect. Another method is to adjust the offset to match the integrated current (restricting the integral to inside the wire) to the known injected current~\cite{Ku2019},  but this is sensitive to how exactly the edges of the wire are defined and may not be a valid technique for certain current carrying objects.

While current reconstruction from different magnetic field sources results in small differences, these deviations from the actual current density can be crucial for measurements with small signal to noise ratio (SNR) or subtle current distribution modifications. To ensure the most reliable result the use of the in-plane magnetic field components is thus preferable. While this is relatively straightforward with widefield imaging~\cite{Simpson2016, Tetienne2017a}, vector magnetometry with high-resolution scanning NV systems has not yet been demonstrated. Our results may motivate the development of scanning probes incorporating multiple NV orientations to enable this.

We note that in the case of NV spins that are very close to the conductor ($< 20$~nm), an anomalous reduction in the in-plane field was recently observed, which is currently not understood~\cite{Tetienne2019}. As this leads to a clearly erroneous current density, in this case it is preferable to use $B_z$. We did not observe such an anomaly in the present experiments where the NV layer had a depth of 120~nm.


\section{Reconstruction of magnetisation}\label{Sec: mag}
 
We now move on to reconstruction of magnetisation and examine whether vector measurements may be beneficial. In order to perform the reconstruction we make the assumptions that the magnetisation is confined to a 2D plane with a vector $\vect{M}(x,y)$ and that the stray field is measured in a parallel plane at a known height $z'$, $\vect{B}(x,y,z')$. The relationship between magnetisation and magnetic fields can be greatly simplified in Fourier-space~\cite{Lima2009,Casola2018,Tan1996,Dreyer2007}, resulting in   
\begin{equation} \label{Eq: mag trans}
\begin{bmatrix} \mathcal{B}_x \\  \mathcal{B}_y \\ \mathcal{B}_z \end{bmatrix} 
=	-\frac{1}{\alpha}
\begin{bmatrix} k_x^2/k & k_x k_y/k &  i k_x \\  k_x k_y/ k & k_y^2/k & i k_y \\  i k_x &  i k_y &  -k \end{bmatrix} 
\begin{bmatrix} \mathcal{M}_x \\  \mathcal{M}_y \\ \mathcal{M}_z  \end{bmatrix} 
\end{equation}
where $\mathcal{B}(k_x,k_y,z')$ and $\mathcal{M}(k_x,k_y)$ are the Fourier-space magnetic field and magnetisation vectors, respectively, and $\alpha$ is the same as in the current reconstruction, $\alpha = 2 e^{k z'} / \mu_0$. 

Since the equations for the three different magnetic field components are linearly dependent, there is effectively only one equation that can be used, e.g. $\mathcal{B}_z=f(\mathcal{M}_x, \mathcal{M}_y, \mathcal{M}_z)$, and so there is an infinite number of solutions for $\mathcal{M}_x$, $\mathcal{M}_y$, and $\mathcal{M}_z$. However, in the case where the direction of the $\textbf{M}$ vector is known and uniform (unit vector ${\bf u_M}$), then we can write $\vect{M}(x,y)= {\bf u_M} M(x,y)$ where $M(x,y)$ is the scalar amplitude. With only one unknown, it is now possible to use a given component of ${\bf B}$, or a combination of components, to deduce $M(x,y)$.

 \begin{figure*}
	\centering
	\includegraphics{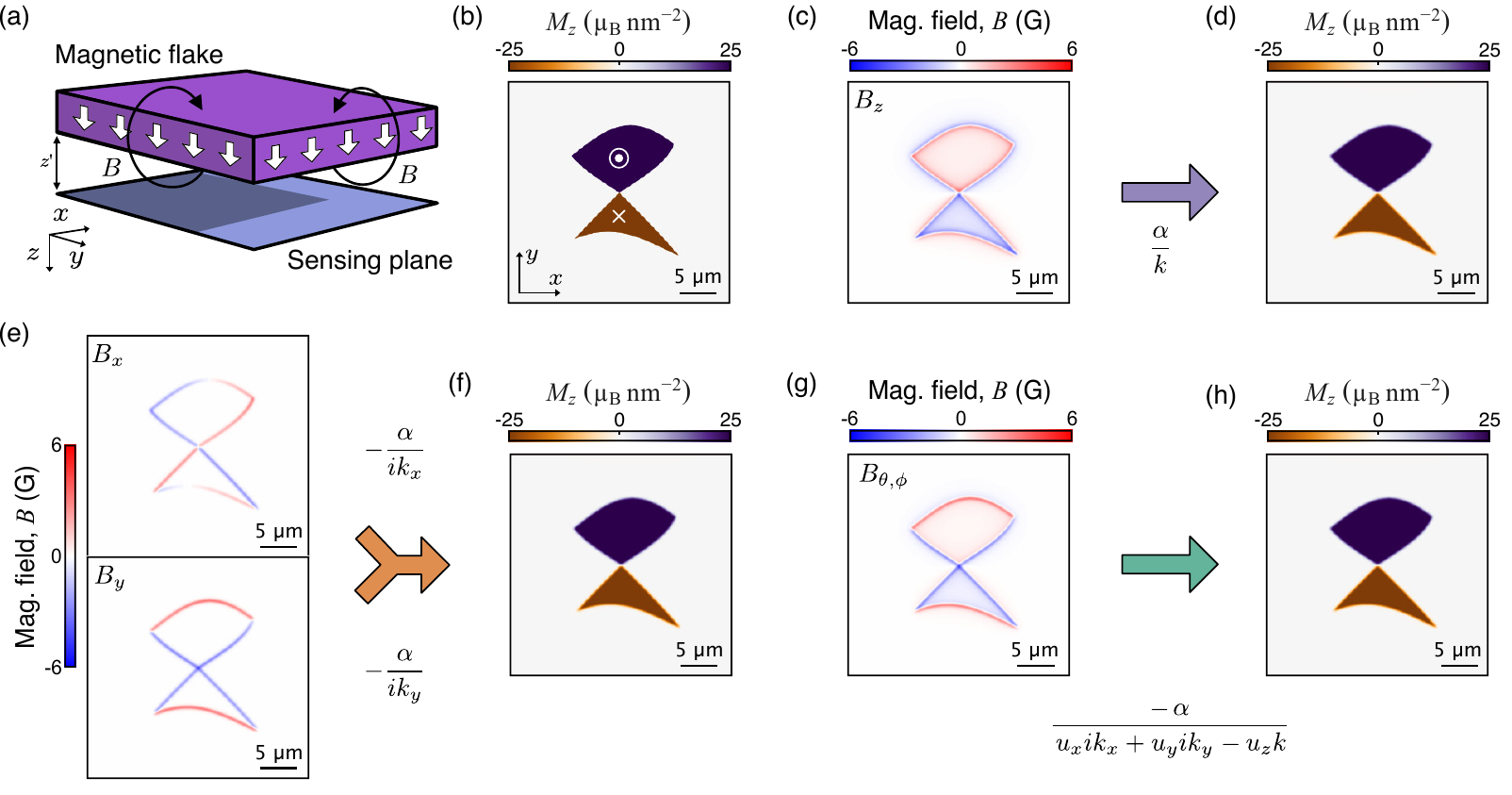}
	\centering
	\caption{\textbf{Reconstruction of out-of-plane magnetisation.} 
		(a) Diagram of out-of-plane magnetised flake with a sensing plane at a height $z'$. 
		(b) Assumed magnetisation map, where two adjacent flakes have magnetisations of same amplitude ($M_z = 25$ \magunit) but opposite signs. 
		(c, d) Calculated $B_z$ map (c) and the reconstruction of $M_z$ (d) using Eq.~\ref{Eq: Bz to Mz}. 
		(e, f) Calculated $B_{x}$ and $B_y$ maps (e) and the reconstruction to $M_z$ (f) using the combined magnetic fields (Eq.~\ref{Eq: Bxy to Mz}). 
		(g, h) Calculated \BNV~map (g) and the reconstruction to $M_z$ (h) using Eq.~\ref{Eq: BNV to Mz}, with the \BNV~magnetic field simulated using \angles{54.7}{0}.
	}
	\label{Fig: Mz intro}
\end{figure*}

From Eq.~\ref{Eq: mag trans}, it is clear that there will be more singularities than in the current reconstruction case. These singularities can be grouped in three types
\begin{enumerate}
	\item $\dfrac{1}{k}$, single point.
	\item $\dfrac{1}{k_{x,y}}$, singularity line.
	\item $\dfrac{1}{k_xk_y}$, two perpendicular singularity lines.
\end{enumerate}
These singularities in Fourier-space correspond to undetermined quantities in real-space. For single point singularities ($1/k$), there is an unknown DC offset, which was also present in the current reconstruction except in the $B_{xy}$ pathway. For lines of singularity ($1/k_{x,y}$ or $1/k_xk_y$) there is an unknown offset for every single line of pixels that are parallel to the line of singularity. For example, with a singularity line of $1/k_x$ each line of pixels in the $x$-direction will have an unknown and different offset.  

In order to form a meaningful real-space image these singularities need to be dealt with such that the gaps of information are closed, i.e. a replacement value for these singularities needs to be picked. For the $1/k$ singularities, just like in the current reconstruction, the condition can be imposed that the real-space magnetisation be null away from the magnetic object. For the lines of singularity, this same condition can be imposed however it needs to be determined for each individual line of pixels. Consequently, this normalisation procedure will add errors in the presence of noise, measurement errors and truncation errors. Thus, it is preferable to use the reconstruction pathway that contains as few singularities as possible. This depends on the direction of ${\bf M}$, and so in what follows we examine different situations, starting with the case of pure $M_z$.

We note that, given the equivalence between current and magnetisation, one could also first reconstruct {\bf J} by inverting Eq.~\ref{Eq: curr transform simp}, generalised so as to include an out-of-plane current component $J_z$, and then deduce {\bf M} by inverting ${\bf J}=\nabla\times{\bf M}$~\cite{Casola2018}. However, this two-step method would introduce the same singularities as the direct inversion of Eq.~\ref{Eq: mag trans}. Thus, for simplicity we perform the direct inversion.

\section{Out-of-plane magnetisation}\label{Sec: out-of-plane}

In the case of magnetic thin films with perpendicular anisotropy, the magnetisation can generally be well approximated by a magnetisation vector $\mathbf{M}(x,y)=M(x,y)\mathbf{u}_z$, where we ignore the small regions where it may lie in the plane (e.g. in domain walls). In this section, we consider this scenario and analyse the different reconstruction pathways. We then compare these findings with experiments performed on flakes of a van der Walls magnetic material, VI$_3$.

\subsection{Theory}\label{Sec: out-of-plane theory}

The transformation for the different magnetic fields components are given by,
\begin{align}
\mathcal{M}_z &= -\frac{ \alpha \mathcal{B}_x}{i k_x}, \\
\mathcal{M}_z &= - \frac{ \alpha \mathcal{B}_y}{i k_y}, \\
\mathcal{M}_z &= \frac{\alpha\mathcal{B}_z}{k}. \label{Eq: Bz to Mz}
\end{align} 
From these equations it can be seen that $\mathcal{B}_z$ has a single-point singularity whereas $\mathcal{B}_x$ and $\mathcal{B}_y$ have a line of singularity. However, in the spirit of current reconstruction, it is interesting to examine the possibility of combining $B_x$ and $B_y$. For instance, we can define a transformation that takes an average of the information afforded by $B_x$ and $B_y$ outside the singularity lines, but use only the singularity-free component otherwise. That is, 
\begin{equation} \label{Eq: Bxy to Mz}
\mathcal{M}_z = - \alpha
\begin{cases} 	
 	\frac{\mathcal{B}_x}{ik_x} & \text{if $k_y =0$}, \\
 	\frac{\mathcal{B}_y}{ik_y} & \text{if $k_x =0$}, \\
	\frac{1}{2} \left(  \frac{\mathcal{B}_y}{ik_y} + \frac{\mathcal{B}_x}{ik_x} \right)  & \text{otherwise}.
\end{cases}
\end{equation} 
By combining the two transverse magnetic field components the transformation for $B_{xy}$ now only experiences a single point singularity ($k=0$) and as such will be the technique used going forward along with the $B_z$ pathway. Additionally, for a measurement along a single arbitrary direction ($\theta,\phi$), the transformation is
 \begin{equation}\label{Eq: BNV to Mz}
 \mathcal{M}_z =  -\frac{ \alpha \FBNV}{u_x i k_x + u_y i k_y - u_z k }, \\
 \end{equation} 
where \unv $= (u_x, u_y, u_z) = (\sin\theta\cos\phi, \sin\theta\sin\phi,\cos\theta)$. In this case, the transformation only has a single point singularity $1/k$ except for the special case where \unv~coincides with the $x$ or $y$ axis, which we will ignore. As all of these pathways ($B_z$, $B_{xy}$, \BNV) result in the same style of singularity they are all normalised in the same manner, that is, we impose the condition that the magnetisation should be null at the edge of the image (zero-point normalisation). 

To test the different reconstruction pathways, we simulated the case of two adjacent flakes with uniform magnetisation of identical amplitude but opposite signs  (Fig.~\ref{Fig: Mz intro}a,b). To simulate conditions that are similar to a widefield imaging experiment, the simulated ${\bf B}$ had a pixel size of 100~nm which is then convolved with a Gaussian with a width of 300~nm. The magnetic field was calculated with an image size of 30~$\mu$m and with a sample to imaging plane stand-off of $z' = 50$~nm (Fig.~\ref{Fig: Mz intro}c,e,g). Before applying the inversion, the ${\bf B}$ maps were padded with zeros on the outside to decrease edge related errors. Unlike the magnetic fields from current, the $B_z$ field from a step in $M_z$ does not have a long range decay and as such truncation is not a significant effect that needs to be considered for widefield experiments. However, in scanning experiments the edge of the image can often cut through a flake and as such has a maximal truncation artefact. 

The magnetisation reconstruction from the $B_z$ magnetic field (Fig.~\ref{Fig: Mz intro}c) transforms without any significant artefacts (Fig.~\ref{Fig: Mz intro}d) as there is no truncation and the zero-point normalisation is straightforward. There is a small deviation at the edge of the flake that is due to the Gaussian convolution of the data set. Using Eq.~\ref{Eq: Bxy to Mz} which combines both transverse magnetic field maps (Fig.~\ref{Fig: Mz intro}e), returns a result that is similar to that of reconstruction from $B_z$ (Fig.~\ref{Fig: Mz intro}f). Likewise, reconstruction from \BNV~(Fig.~\ref{Fig: Mz intro}g) also returns a reliable magnetisation (Fig.~\ref{Fig: Mz intro}h). 

\begin{figure}
	\centering
	\includegraphics[width=\columnwidth]{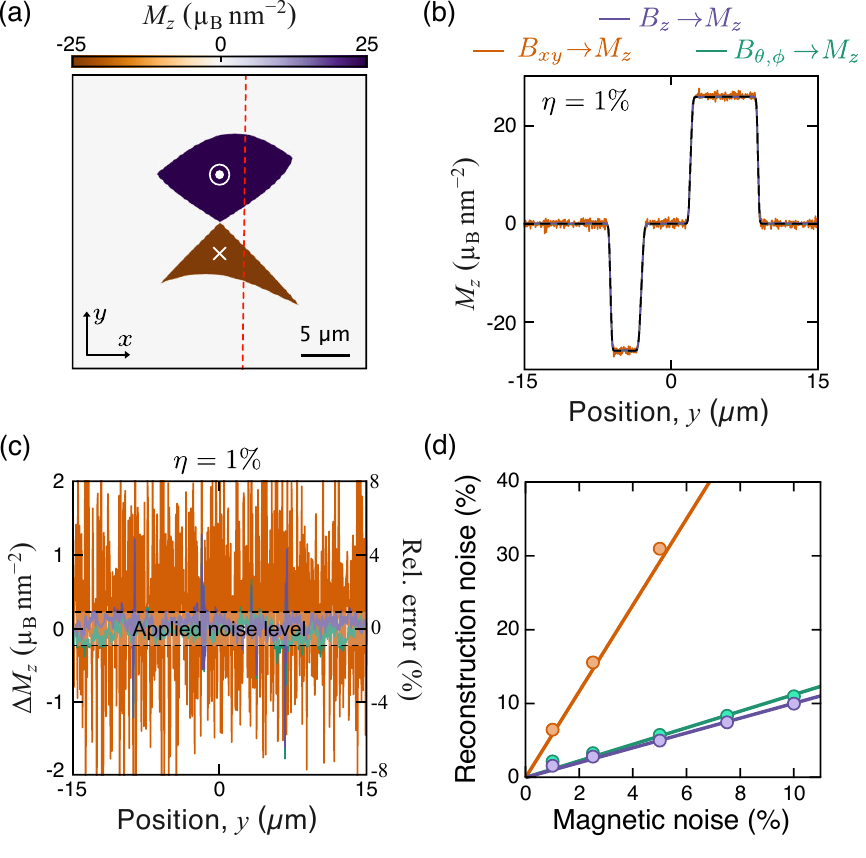}
	\caption{\textbf{Effect of noise on reconstruction of out-of-plane magnetisation.} 
		(a) Simulated magnetisation map, identical to Fig.~\ref{Fig: Mz intro}b.
		(b) Vertical line cut along the red dashed line in (a) for magnetisation reconstruction from different magnetic field components (solid lines) and the simulated magnetisation (dashed line) with an applied noise of $\eta = 1$\%. The $B_z$ and \BNV~cases are barely visible here but can be better seen in (c).
		(c) Line cuts of the difference between the simulated and reconstructed magnetisation for different magnetic components. The standard deviation of the applied noise is shown as a shaded region between two dashed lines.  
		(d) Simulation of the transformation of the magnetic noise to magnetisation noise for different magnetic field components. Linear fits to the data is also shown under the assumption that the two noises converge at zero. 
		The \BNV~magnetic field was simulated using \angles{54.7}{0}.
	}
	\label{Fig: Mz noise}
\end{figure}

\begin{figure*}
	\centering
	\includegraphics{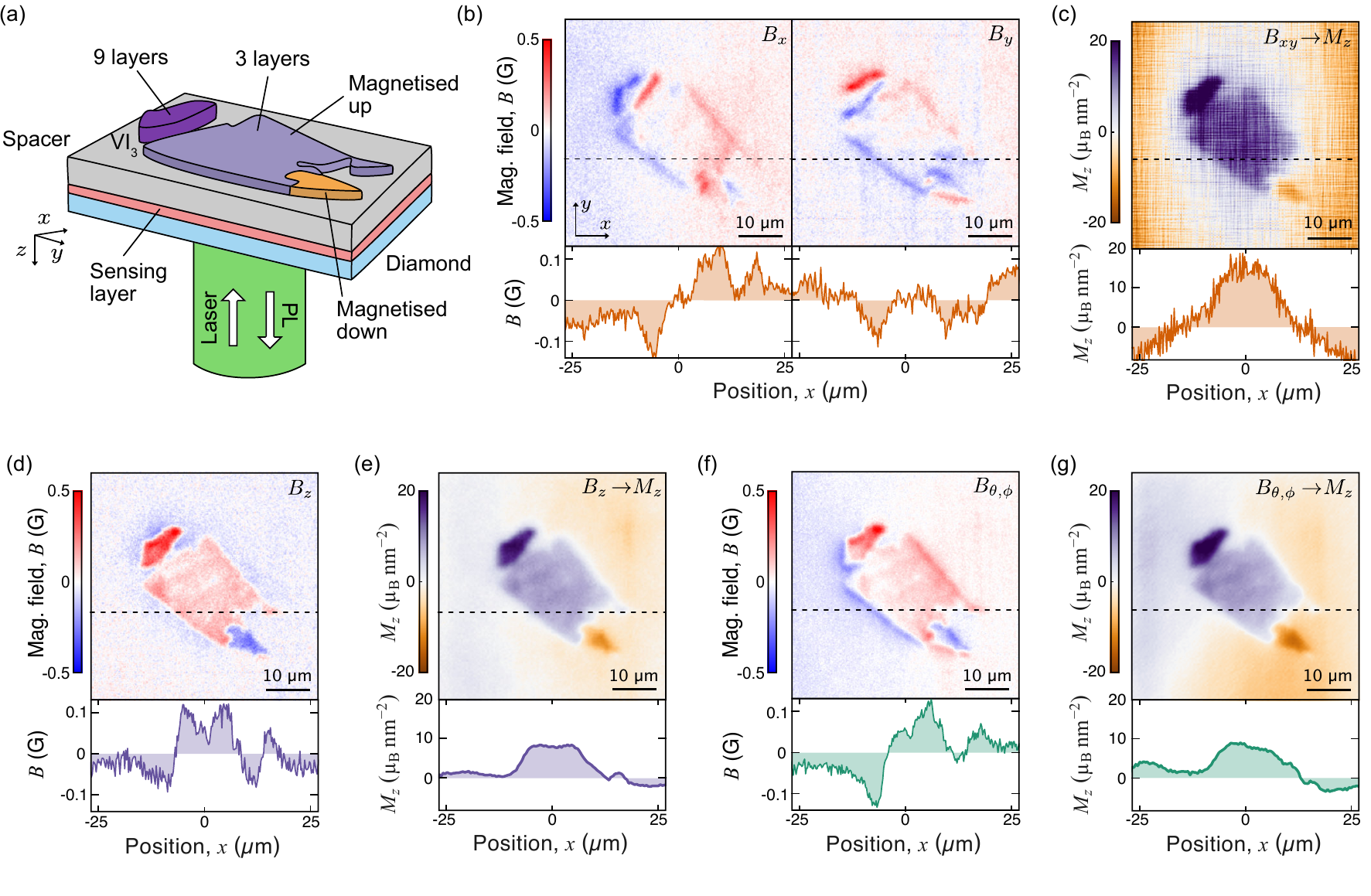}
	\caption{\textbf{Experimental example of reconstruction of out-of-plane magnetisation.} 
		(a) Schematic of experimental setup where magnetic VI$_3$ flakes are placed onto the diamond surface. The optical excitation and photoluminescence (PL) readout is performed through the diamond substrate. 
		(b) Transverse magnetic field components, $B_x$ and $B_y$, obtained from full vector ODMR spectroscopy. 
		(c) Reconstruction of out-of-plane magnetisation from the transverse magnetic fields. 
		(d) $B_z$ magnetic field obtained from the same full vector ODMR measurement as in (b).  
		(e) Reconstruction of out-of-plane magnetisation from the $B_z$ magnetic field. 
		(f) \BNV~from a single NV axis with a direction of \angles{54.7}{45}.
		(g) Reconstruction of out-of-plane magnetisation from the \BNV~magnetic field. 
		In (b-g), below each map is a line cut taken along the dashed line shown on the map. 
	}
	\label{Fig: Mz exp}
\end{figure*}

\subsection{Noise propagation}

Previous work on different reconstruction methods for current density have discussed in detail how noise can affect the reconstruction process~\cite{Meltzer2017,Clement2019}. To elucidate the differences between the magnetisation reconstruction from different magnetic field components, we apply noise to the magnetic field before reconstruction and take vertical line cuts across the flake (Fig.~\ref{Fig: Mz noise}a) for all the different reconstruction pathways (Fig.~\ref{Fig: Mz noise}b). Here we simulate random noise to each pixel in the magnetic field maps, characterised by a standard deviation $\eta$ defined relative to the maximum field amplitude in the image. Additionally, we remove the Gaussian convolution to compare directly the effect of noise on the transformation without additional treatment. With the introduction of a relatively small error of $\eta = 1\%$ there is a drastic change in the quality of the reconstruction for different magnetic field components  (Fig.~\ref{Fig: Mz noise}b). When compared to the simulated flake (black dashed line) the reconstructions from $B_z$ and \BNV~perform well, however, reconstruction from $B_{xy}$ has significantly more noise. This noise amplification is quantified by taking the difference between the simulated ($M_z^s$) and reconstructed ($M_z^p$) magnetisation, $\Delta M_z = M_z^s - M_z^p$ normalised to the assumed magnetisation amplitude (Fig.~\ref{Fig: Mz noise}c). The error in the reconstructed magnetisation shows that reconstruction from $B_{xy}$ is the least robust to noise, taking the initial noise of 1\% and returning noise $> 4\%$. In contrast, the other reconstruction pathways have a significantly better response and maintain a similar noise level. In principle, techniques can be used to minimise the error on the reconstruction from $B_{xy}$, as will be used for in-plane magnetisation, but they come at the cost of spatial resolution. 

The transformation of the magnetic field noise to magnetisation noise has a linear response (Fig.~\ref{Fig: Mz noise}d), where reconstruction from both $B_z$ and \BNV~fields closely mirrors the applied noise, while $B_{xy}$ drastically amplifies it. The difference in the reconstruction is partially due to the transverse magnetic field maps being more sparsely filled than $B_z$ which results in a difference in the effective SNR after transformation. It is important to note that in an experiment with (100)-orientated diamond, $B_z$ is obtained through full vector ODMR spectroscopy which has a worse SNR than a single NV measurement, and as such may in practice perform worse. Lastly, in the case of truncated fields, the symmetry of the $B_z$ magnetic field map translates to a symmetric erroneous background magnetisation, whereas due to the orientation of \BNV~the background magnetic field is asymmetric. Both forms of background magnetisations can be removed through an appropriate background subtraction, but different approaches may be required as was the case in current reconstruction.

For both ensemble NV imaging~\cite{Tetienne2017a} and scanning NV microscopy~\cite{Maletinsky2012} reconstruction from $B_z$ offers the best transformation potential. As a consequence, (111)-orientated diamond imaging platforms offer both the best transformation and the best signal to noise ratio by not requiring full vector magnetometry. Bulk diamond slabs with (111)-orientation are commercially available and recently (111)-diamond atomic force microscope (AFM) scanning tips have been fabricated~\cite{Rohner2019}. However, \BNV~does perform very similarly in (100)-orientated diamond and is appropriate for most sensing scenarios. 

\subsection{Experimental comparison}\label{Sec: exp out-of-plane}

\begin{figure*}
	\centering
	\includegraphics{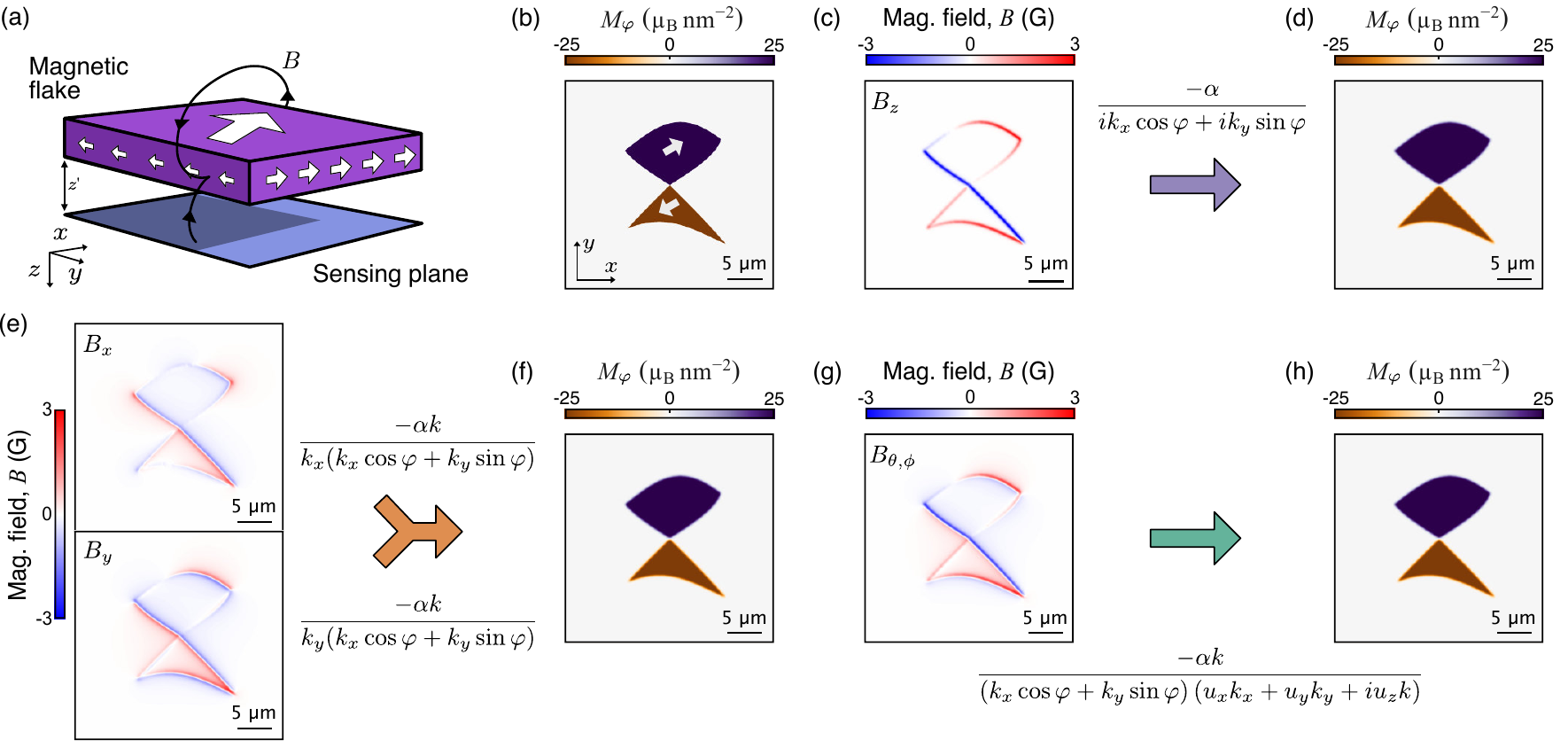}
	\caption{\textbf{Reconstruction of in-plane magnetisation.} 
		(a) Diagram of magnetic flake with in-plane magnetisation at $\varphi = 30^\circ$ with a sensing plane. 
		(b) Assumed magnetisation map, corresponding to two adjacent flakes with opposite magnetisation as denoted by the arrows. 
		(c, d) Calculated $B_z$ map (c) and the reconstruction of $M_\varphi$ (d) using Eq.~\ref{Eq: Bz to mtheta}. 
		(e, f) Calculated $B_{x}$ and $B_y$ maps (e) and the reconstruction to $M_\varphi$ (f) using the combined magnetic fields (Eq.~\ref{Eq: Bxy to mtheta}). 
		(g, h) Calculated \BNV~map (g) and the reconstruction to $M_\varphi$ (h) using Eq.~\ref{Eq: bnv to mtheta}, with \angles{54.7}{45}. 
	}
	\label{Fig: M theta intro}
\end{figure*}

Experimentally, we validate the simulations using a flake of VI$_3$~\cite{Tian2019,Son2019,Kong2019} placed onto the diamond (Fig.~\ref{Fig: Mz exp}a), where the description of the fabrication process can be found in Ref.~\cite{Broadway2020}. The measurements are taken using an NV layer similar to that used for current reconstruction in Sec.~\ref{Sec: exp curr}. The measurements were performed at a temperature of $T = 5$~K under a bias field of $B = 120$~G to distinctly split all four NV orientations and were taken using a custom-built closed-cycle cryostat widefield florescence microscope~\cite{Lillie2020}. The magnetic field components were obtained in the same fashion as in the measurements of the Nb wire (Sec.~\ref{Sec: exp curr}).  

The magnetic field maps are shown in Fig.~\ref{Fig: Mz exp}b, d, f, where the bias field was removed by subtracting the mean value of each map. The corresponding reconstructed magnetisation $M_z$ maps are shown in Fig.~\ref{Fig: Mz exp}c, e, g. Due to the transformation for the transverse magnetic fields amplifying the noise, the reconstruction from this pathway is littered with additional artefacts (Fig.~\ref{Fig: Mz exp}c). In contrast, the reconstruction from the $B_z$ magnetic field produces a relatively uniform near zero background and a clear domain structure in the flake  (Fig.~\ref{Fig: Mz exp}e). This is also in contrast to the reconstruction from the \BNV~magnetic field (Fig.~\ref{Fig: Mz exp}g) taken with an NV with an orientation of \angles{54.7}{45}. The \BNV~magnetic field has a gradient in the measurement (presumably an artefact arising from the ODMR fitting) that when reconstructing to magnetisation generates a (non-physical) background magnetisation that hinders quantitative analysis of the magnetisation inside the flake. Although there are strategies to remove this background, $B_z$ seems to offer a more reliable reconstruction, particularly in cases where there are additional nearby flakes, whose interfering magnetic field may inhibit such techniques. 

Here we have found that without additional processing, only reconstruction of out-of-plane magnetisation using the $B_z$ magnetic field is able to reconstruct the magnetisation of the flake such that there are clearly distinguishable magnetic domains with relatively uniform magnetisation in each. Our results thus confirm that $B_z$ is the best choice for reconstructing out-of-plane magnetisation if available or reconstruction from \BNV~if additional stray background magnetic fields are minimised.

\section{In-plane magnetisation}\label{Sec: in plane}

We now move to the case of an in-plane magnetisation. Here the magnetisation can be reconstructed from the stray magnetic field only for a system with in-plane uniaxial anisotropy so that the magnetisation vector has a fixed and known direction throughout the sample. We analyse the different pathways to perform this reconstruction and discuss the singularities involved and how they are affected by noise.

The in-plane magnetisation has a more complex transformation from magnetic field as can be seen from the total transformation given in Eq.~\ref{Eq: mag trans}. The magnetisation vector can be expressed as $\vect{M}=M_\varphi (\cos(\varphi),\sin(\varphi),0)$ where $M_\varphi$ is the amplitude and $\varphi$ is the angle of $\vect{M}$ with respect to the $x$-axis. The transformations take the form of
\begin{align}
	\mathcal{M}_\varphi^x &= - \frac{\alpha k\mathcal{B}_x}{ k_x(k_x\cos\varphi + k_y \sin\varphi) }, \\
	\mathcal{M}_\varphi^y &= -  \frac{ \alpha k\mathcal{B}_y}{ k_y(k_x\cos\varphi + k_y \sin\varphi) }, \\
	\mathcal{M}_\varphi^z &= -\frac{ \alpha \mathcal{B}_z}{ ik_x\cos\varphi + ik_y \sin\varphi }. \label{Eq: Bz to mtheta}
\end{align}
These transformations all exhibit a singularity line that is related to the angle of magnetisation, i.e. when $k_x = -k_y \tan\varphi$. The transverse magnetic field transformations also have an additional singularity line when $k_x = 0$ for $B_x$ and when $k_y = 0$ for $B_y$. This additional singularity is circumvented in the same way as in the out-of-plane magnetisation by combining the two cases, 
\begin{equation}\label{Eq: Bxy to mtheta}
\mathcal{M}_\varphi^{x,y} = 
\begin{cases} 	
\mathcal{M}_\varphi^x  & \text{if $k_y =0$}, \\
\mathcal{M}_\varphi^y & \text{if $k_x =0$}, \\
\frac{1}{2} \left(  \mathcal{M}_\varphi^x + \mathcal{M}_\varphi^y \right)  & \text{otherwise},
\end{cases}
\end{equation}
returning the transformation to the same singularities as $B_z$. To bypass the singularities we set each point along this singularity line by imposing that each corresponding oblique linecut in the real-space $M_\varphi$ map have a vanishing baseline, based on the fact that there should be no magnetisation outside the flake. Therefore, the presence of noise and errors in the $B$ maps (e.g. background gradients) will translate into an error in each of these offsets, unlike $M_z$ where only the global offset is uncertain. Reconstruction from \BNV~can also be used and takes the more complex form, 
\begin{equation}
\mathcal{M}_\varphi^{\theta,\phi} = \frac{ - \alpha k\FBNV}{ \left(k_x \cos\varphi + k_y \sin\varphi\right) \left( u_x k_x + u_y k_y + i u_z k \right)},  \label{Eq: bnv to mtheta}
\end{equation} 
which has the same singularities ($k_x = -k_y \tan\varphi$) as the other transformations.

\begin{figure}
	\centering
	\includegraphics[width=\columnwidth]{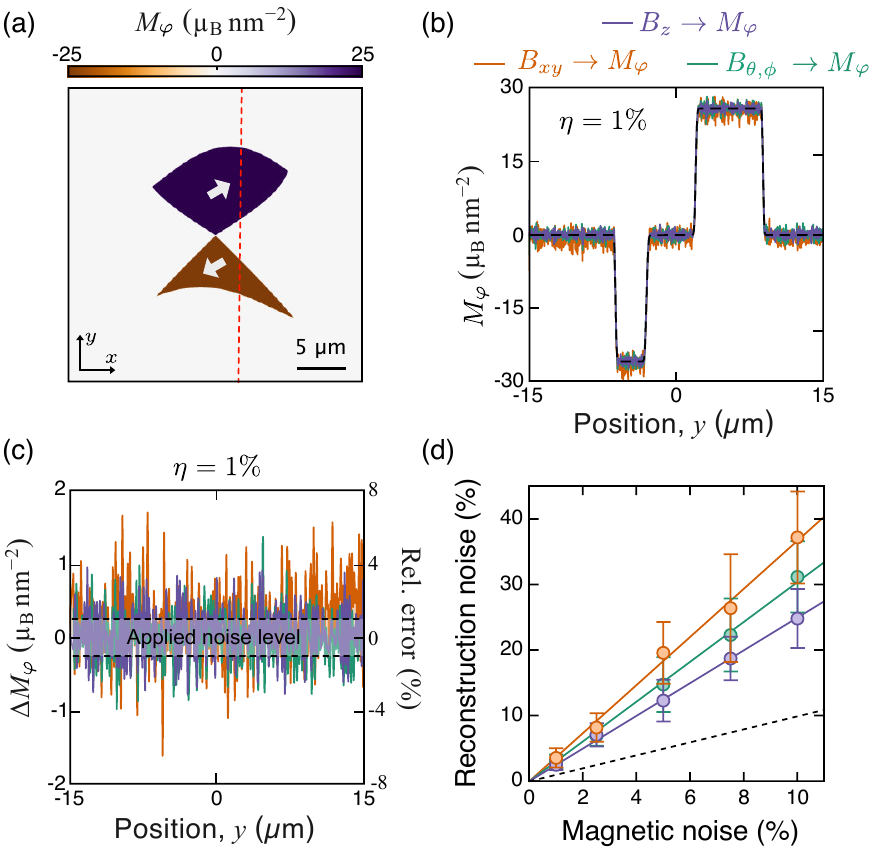}
	\caption{ \textbf{Effect of noise on reconstruction of in-plane magnetisation.} 
		(a) Simulated magnetisation map, identical to Fig.~\ref{Fig: M theta intro}b with an angle of $\varphi = 30^\circ$.
		(b) Vertical line cut along the red dashed line in (a) for magnetisation reconstruction from different magnetic field components with a relative noise $\eta = 1$\%  (solid lines) and the simulated magnetisation (dashed line). The $B_z$ and \BNV cases are barely visible here but can be better seen in (c).
		(c) Line cuts of the difference between the simulated and reconstructed magnetisation for different magnetic components. The standard deviation of the applied field noise is shown as a shaded region between two dashed lines.  
		(d) Simulation of the transformation of the magnetic field noise to magnetisation noise for different magnetic field components where the error bars represent the standard deviation in the reconstructed noise over multiple different simulations. The magnetic field noise level is shown as a black dashed line and linear fits to the data is also shown under the assumption that the two noises converge at zero. 
		The \BNV~magnetic field was simulated using \angles{54.7}{0}.		
	}
	\label{Fig: in-plane mag}
\end{figure}

The reconstruction methods were tested by simulating considering two oppositely magnetised flakes in close proximity with an angle of $\varphi = 30^\circ$ from the $x$-axis (Fig.~\ref{Fig: M theta intro}a,b), where it was assumed that we knew the angle of the magnetisation for the reconstruction. In principle, for a single domain it is possible to determine the angle through reconstruction from different angles and excluding non physical results. The different reconstruction pathways in the absence of noise all perform similarly, where reconstruction from $B_z$ (Fig.~\ref{Fig: M theta intro}c,d), $B_{xy}$ (Fig.~\ref{Fig: M theta intro}e,f), and \BNV~(Fig.~\ref{Fig: M theta intro}g,h) all perform well with a blurring of the flake edge due to the imposed diffraction limit of $300~$nm.

To illustrate the effect of noise we take vertical line cuts across the flakes ( Fig.~\ref{Fig: in-plane mag}a). The inclusion of a small random noise $\eta = 1\%$  in the magnetic field images results in a similar response across all pathways (Fig.~\ref{Fig: in-plane mag}b). However, unlike in the case of out-of-plane magnetisation here the noise in the magnetisation is amplified compared to the magnetic field noise, by a factor 2-3 (Fig.~\ref{Fig: in-plane mag}c). The linear scaling of the noise from the different reconstructions indicates that the $B_z$ transformation is the most robust, although it still amplifies the noise by more than a factor of 2 (Fig.~\ref{Fig: in-plane mag}d). We note that the noise amplification greatly depends on the type of filter one applies to the transformation. In this case we apply an Hanning filter with a low pass filter to remove oscillations that are faster than $1/z'$. However, one could apply stricter filters at the loss of spatial resolution to reduce the noise further. These results suggest that reconstruction from the $B_z$ magnetic field component has an advantage to other components due to it being slightly more robust to noise. This will be particularly relevant for scenarios with a low signal to noise or subtle edge effects. 

\section{conclusion}

In this work we investigated how 2D maps of stray magnetic fields can be used to reconstruct the source quantity, focusing on the comparison between the different pathways afforded by vector information. Additionally, we investigated the same reconstructions with an arbitrary projection of the magnetic field (\BNV). In the case of reconstruction of current density, we find that the combination of $B_x$ and $B_y$ magnetic field maps provides a near-ideal transformation, while $B_z$ and \BNV~measurements lead to edge artefacts due to truncation of the long range $B_z$ magnetic field, which is particularly prone to error for small image sizes which are commonly used in scanning experiments. These results were confirmed experimentally by performing the reconstruction from different magnetic field components from a current carrying wire with a widefield NV microscope. These findings motivate the development of NV scanning probes with vector capability, which would enable high spatial resolution, high accuracy imaging of transport processes.

In the case of magnetisation reconstruction, we find that using the combination of $B_{x}$ and $B_y$ leads to a significant amplification of noise. As a consequence this reconstruction pathway is undesirable when compared with $B_z$ and \BNV, which both produce a relatively similar quality for reconstruction of magnetisation. Independent of the magnetisation direction, we find that reconstruction from the $B_z$ magnetic field component is the most robust to noise and thus the preferred reconstruction pathway. We confirm these results through performing magnetisation reconstruction on perpendicularly-magnetised flakes using a widefield NV microscope. These findings motivate the use of (111)-oriented diamond with out-of-plane orientated NVs for magnetisation mapping~\cite{Rohner2019}.\\

\section{Acknowledgements}

The authors thank B. C. Johnson for assistance with the preparation of diamond samples, C. Tan, G. Zheng, L. Wang, S. Tian, C. Li and H. Lei for providing the VI$_3$ samples (described in Ref.~\cite{Broadway2020}) and L. Thiel for useful discussions. We acknowledge support from the Australian Research Council (ARC) through grants DE170100129, CE170100012, LE180100037 and DP190101506. D.A.B. and S.E.L. are supported by an Australian Government Research Training Program Scholarship. 

\bibliographystyle{apsrev4-1}
\bibliography{bib}

\begin{thebibliography}{57}%
\makeatletter
\providecommand \@ifxundefined [1]{%
 \@ifx{#1\undefined}
}%
\providecommand \@ifnum [1]{%
 \ifnum #1\expandafter \@firstoftwo
 \else \expandafter \@secondoftwo
 \fi
}%
\providecommand \@ifx [1]{%
 \ifx #1\expandafter \@firstoftwo
 \else \expandafter \@secondoftwo
 \fi
}%
\providecommand \natexlab [1]{#1}%
\providecommand \enquote  [1]{``#1''}%
\providecommand \bibnamefont  [1]{#1}%
\providecommand \bibfnamefont [1]{#1}%
\providecommand \citenamefont [1]{#1}%
\providecommand \href@noop [0]{\@secondoftwo}%
\providecommand \href [0]{\begingroup \@sanitize@url \@href}%
\providecommand \@href[1]{\@@startlink{#1}\@@href}%
\providecommand \@@href[1]{\endgroup#1\@@endlink}%
\providecommand \@sanitize@url [0]{\catcode `\\12\catcode `\$12\catcode
  `\&12\catcode `\#12\catcode `\^12\catcode `\_12\catcode `\%12\relax}%
\providecommand \@@startlink[1]{}%
\providecommand \@@endlink[0]{}%
\providecommand \url  [0]{\begingroup\@sanitize@url \@url }%
\providecommand \@url [1]{\endgroup\@href {#1}{\urlprefix }}%
\providecommand \urlprefix  [0]{URL }%
\providecommand \Eprint [0]{\href }%
\providecommand \doibase [0]{http://dx.doi.org/}%
\providecommand \selectlanguage [0]{\@gobble}%
\providecommand \bibinfo  [0]{\@secondoftwo}%
\providecommand \bibfield  [0]{\@secondoftwo}%
\providecommand \translation [1]{[#1]}%
\providecommand \BibitemOpen [0]{}%
\providecommand \bibitemStop [0]{}%
\providecommand \bibitemNoStop [0]{.\EOS\space}%
\providecommand \EOS [0]{\spacefactor3000\relax}%
\providecommand \BibitemShut  [1]{\csname bibitem#1\endcsname}%
\let\auto@bib@innerbib\@empty
\bibitem [{\citenamefont {Shi}\ \emph {et~al.}(2019)\citenamefont {Shi},
  \citenamefont {Kahn}, \citenamefont {Niu}, \citenamefont {Fei}, \citenamefont
  {Sun}, \citenamefont {Cai}, \citenamefont {Francisco}, \citenamefont {Wu},
  \citenamefont {Shen}, \citenamefont {Xu}, \citenamefont {Cobden},\ and\
  \citenamefont {Cui}}]{Shi2019}%
  \BibitemOpen
  \bibfield  {author} {\bibinfo {author} {\bibfnamefont {Y.}~\bibnamefont
  {Shi}}, \bibinfo {author} {\bibfnamefont {J.}~\bibnamefont {Kahn}}, \bibinfo
  {author} {\bibfnamefont {B.}~\bibnamefont {Niu}}, \bibinfo {author}
  {\bibfnamefont {Z.}~\bibnamefont {Fei}}, \bibinfo {author} {\bibfnamefont
  {B.}~\bibnamefont {Sun}}, \bibinfo {author} {\bibfnamefont {X.}~\bibnamefont
  {Cai}}, \bibinfo {author} {\bibfnamefont {B.~A.}\ \bibnamefont {Francisco}},
  \bibinfo {author} {\bibfnamefont {D.}~\bibnamefont {Wu}}, \bibinfo {author}
  {\bibfnamefont {Z.-X.}\ \bibnamefont {Shen}}, \bibinfo {author}
  {\bibfnamefont {X.}~\bibnamefont {Xu}}, \bibinfo {author} {\bibfnamefont
  {D.~H.}\ \bibnamefont {Cobden}}, \ and\ \bibinfo {author} {\bibfnamefont
  {Y.-T.}\ \bibnamefont {Cui}},\ }\href {\doibase 10.1126/sciadv.aat8799}
  {\bibfield  {journal} {\bibinfo  {journal} {Sci. Adv.}\ }\textbf {\bibinfo
  {volume} {5}},\ \bibinfo {pages} {eaat8799} (\bibinfo {year}
  {2019})}\BibitemShut {NoStop}%
\bibitem [{\citenamefont {Vasyukov}\ \emph {et~al.}(2013)\citenamefont
  {Vasyukov}, \citenamefont {Anahory}, \citenamefont {Embon}, \citenamefont
  {Halbertal}, \citenamefont {Cuppens}, \citenamefont {Neeman}, \citenamefont
  {Finkler}, \citenamefont {Segev}, \citenamefont {Myasoedov}, \citenamefont
  {Rappaport}, \citenamefont {Huber},\ and\ \citenamefont
  {Zeldov}}]{Vasyukov2013}%
  \BibitemOpen
  \bibfield  {author} {\bibinfo {author} {\bibfnamefont {D.}~\bibnamefont
  {Vasyukov}}, \bibinfo {author} {\bibfnamefont {Y.}~\bibnamefont {Anahory}},
  \bibinfo {author} {\bibfnamefont {L.}~\bibnamefont {Embon}}, \bibinfo
  {author} {\bibfnamefont {D.}~\bibnamefont {Halbertal}}, \bibinfo {author}
  {\bibfnamefont {J.}~\bibnamefont {Cuppens}}, \bibinfo {author} {\bibfnamefont
  {L.}~\bibnamefont {Neeman}}, \bibinfo {author} {\bibfnamefont
  {A.}~\bibnamefont {Finkler}}, \bibinfo {author} {\bibfnamefont
  {Y.}~\bibnamefont {Segev}}, \bibinfo {author} {\bibfnamefont
  {Y.}~\bibnamefont {Myasoedov}}, \bibinfo {author} {\bibfnamefont {M.~L.}\
  \bibnamefont {Rappaport}}, \bibinfo {author} {\bibfnamefont {M.~E.}\
  \bibnamefont {Huber}}, \ and\ \bibinfo {author} {\bibfnamefont
  {E.}~\bibnamefont {Zeldov}},\ }\href {\doibase 10.1038/nnano.2013.169}
  {\bibfield  {journal} {\bibinfo  {journal} {Nat. Nanotechnol.}\ }\textbf
  {\bibinfo {volume} {8}},\ \bibinfo {pages} {639} (\bibinfo {year}
  {2013})}\BibitemShut {NoStop}%
\bibitem [{\citenamefont {Kalisky}\ \emph {et~al.}(2013)\citenamefont
  {Kalisky}, \citenamefont {Spanton}, \citenamefont {Noad}, \citenamefont
  {Kirtley}, \citenamefont {Nowack}, \citenamefont {Bell}, \citenamefont
  {Sato}, \citenamefont {Hosoda}, \citenamefont {Xie}, \citenamefont {Hikita},
  \citenamefont {Woltmann}, \citenamefont {Pfanzelt}, \citenamefont {Jany},
  \citenamefont {Richter}, \citenamefont {Hwang}, \citenamefont {Mannhart},\
  and\ \citenamefont {Moler}}]{Kalisky2013}%
  \BibitemOpen
  \bibfield  {author} {\bibinfo {author} {\bibfnamefont {B.}~\bibnamefont
  {Kalisky}}, \bibinfo {author} {\bibfnamefont {E.~M.}\ \bibnamefont
  {Spanton}}, \bibinfo {author} {\bibfnamefont {H.}~\bibnamefont {Noad}},
  \bibinfo {author} {\bibfnamefont {J.~R.}\ \bibnamefont {Kirtley}}, \bibinfo
  {author} {\bibfnamefont {K.~C.}\ \bibnamefont {Nowack}}, \bibinfo {author}
  {\bibfnamefont {C.}~\bibnamefont {Bell}}, \bibinfo {author} {\bibfnamefont
  {H.~K.}\ \bibnamefont {Sato}}, \bibinfo {author} {\bibfnamefont
  {M.}~\bibnamefont {Hosoda}}, \bibinfo {author} {\bibfnamefont
  {Y.}~\bibnamefont {Xie}}, \bibinfo {author} {\bibfnamefont {Y.}~\bibnamefont
  {Hikita}}, \bibinfo {author} {\bibfnamefont {C.}~\bibnamefont {Woltmann}},
  \bibinfo {author} {\bibfnamefont {G.}~\bibnamefont {Pfanzelt}}, \bibinfo
  {author} {\bibfnamefont {R.}~\bibnamefont {Jany}}, \bibinfo {author}
  {\bibfnamefont {C.}~\bibnamefont {Richter}}, \bibinfo {author} {\bibfnamefont
  {H.~Y.}\ \bibnamefont {Hwang}}, \bibinfo {author} {\bibfnamefont
  {J.}~\bibnamefont {Mannhart}}, \ and\ \bibinfo {author} {\bibfnamefont
  {K.~A.}\ \bibnamefont {Moler}},\ }\href {\doibase 10.1038/nmat3753}
  {\bibfield  {journal} {\bibinfo  {journal} {Nat. Mater.}\ }\textbf {\bibinfo
  {volume} {12}},\ \bibinfo {pages} {1091} (\bibinfo {year}
  {2013})}\BibitemShut {NoStop}%
\bibitem [{\citenamefont {Nowack}\ \emph {et~al.}(2013)\citenamefont {Nowack},
  \citenamefont {Spanton}, \citenamefont {Baenninger}, \citenamefont
  {K{\"{o}}nig}, \citenamefont {Kirtley}, \citenamefont {Kalisky},
  \citenamefont {Ames}, \citenamefont {Leubner}, \citenamefont {Br{\"{u}}ne},
  \citenamefont {Buhmann}, \citenamefont {Molenkamp}, \citenamefont
  {Goldhaber-Gordon},\ and\ \citenamefont {Moler}}]{Nowack2013}%
  \BibitemOpen
  \bibfield  {author} {\bibinfo {author} {\bibfnamefont {K.~C.}\ \bibnamefont
  {Nowack}}, \bibinfo {author} {\bibfnamefont {E.~M.}\ \bibnamefont {Spanton}},
  \bibinfo {author} {\bibfnamefont {M.}~\bibnamefont {Baenninger}}, \bibinfo
  {author} {\bibfnamefont {M.}~\bibnamefont {K{\"{o}}nig}}, \bibinfo {author}
  {\bibfnamefont {J.~R.}\ \bibnamefont {Kirtley}}, \bibinfo {author}
  {\bibfnamefont {B.}~\bibnamefont {Kalisky}}, \bibinfo {author} {\bibfnamefont
  {C.}~\bibnamefont {Ames}}, \bibinfo {author} {\bibfnamefont {P.}~\bibnamefont
  {Leubner}}, \bibinfo {author} {\bibfnamefont {C.}~\bibnamefont
  {Br{\"{u}}ne}}, \bibinfo {author} {\bibfnamefont {H.}~\bibnamefont
  {Buhmann}}, \bibinfo {author} {\bibfnamefont {L.~W.}\ \bibnamefont
  {Molenkamp}}, \bibinfo {author} {\bibfnamefont {D.}~\bibnamefont
  {Goldhaber-Gordon}}, \ and\ \bibinfo {author} {\bibfnamefont {K.~A.}\
  \bibnamefont {Moler}},\ }\href {\doibase 10.1038/nmat3682} {\bibfield
  {journal} {\bibinfo  {journal} {Nat. Mater.}\ }\textbf {\bibinfo {volume}
  {12}},\ \bibinfo {pages} {787} (\bibinfo {year} {2013})}\BibitemShut
  {NoStop}%
\bibitem [{\citenamefont {Dinner}\ \emph
  {et~al.}(2007{\natexlab{a}})\citenamefont {Dinner}, \citenamefont {Moler},
  \citenamefont {Feldmann},\ and\ \citenamefont {Beasley}}]{Dinner2007}%
  \BibitemOpen
  \bibfield  {author} {\bibinfo {author} {\bibfnamefont {R.~B.}\ \bibnamefont
  {Dinner}}, \bibinfo {author} {\bibfnamefont {K.~A.}\ \bibnamefont {Moler}},
  \bibinfo {author} {\bibfnamefont {D.~M.}\ \bibnamefont {Feldmann}}, \ and\
  \bibinfo {author} {\bibfnamefont {M.~R.}\ \bibnamefont {Beasley}},\ }\href
  {\doibase 10.1103/PhysRevB.75.144503} {\bibfield  {journal} {\bibinfo
  {journal} {Phys. Rev. B}\ }\textbf {\bibinfo {volume} {75}},\ \bibinfo
  {pages} {144503} (\bibinfo {year} {2007}{\natexlab{a}})}\BibitemShut
  {NoStop}%
\bibitem [{\citenamefont {Dinner}\ \emph
  {et~al.}(2007{\natexlab{b}})\citenamefont {Dinner}, \citenamefont {Moler},
  \citenamefont {Beasley},\ and\ \citenamefont {Feldmann}}]{Dinner2007a}%
  \BibitemOpen
  \bibfield  {author} {\bibinfo {author} {\bibfnamefont {R.~B.}\ \bibnamefont
  {Dinner}}, \bibinfo {author} {\bibfnamefont {K.~A.}\ \bibnamefont {Moler}},
  \bibinfo {author} {\bibfnamefont {M.~R.}\ \bibnamefont {Beasley}}, \ and\
  \bibinfo {author} {\bibfnamefont {D.~M.}\ \bibnamefont {Feldmann}},\ }\href
  {\doibase 10.1063/1.2740610} {\bibfield  {journal} {\bibinfo  {journal}
  {Appl. Phys. Lett.}\ }\textbf {\bibinfo {volume} {90}},\ \bibinfo {pages}
  {212501} (\bibinfo {year} {2007}{\natexlab{b}})}\BibitemShut {NoStop}%
\bibitem [{\citenamefont {Doherty}\ \emph {et~al.}(2013)\citenamefont
  {Doherty}, \citenamefont {Manson}, \citenamefont {Delaney}, \citenamefont
  {Jelezko}, \citenamefont {Wrachtrup},\ and\ \citenamefont
  {Hollenberg}}]{Doherty2013}%
  \BibitemOpen
  \bibfield  {author} {\bibinfo {author} {\bibfnamefont {M.~W.}\ \bibnamefont
  {Doherty}}, \bibinfo {author} {\bibfnamefont {N.~B.}\ \bibnamefont {Manson}},
  \bibinfo {author} {\bibfnamefont {P.}~\bibnamefont {Delaney}}, \bibinfo
  {author} {\bibfnamefont {F.}~\bibnamefont {Jelezko}}, \bibinfo {author}
  {\bibfnamefont {J.}~\bibnamefont {Wrachtrup}}, \ and\ \bibinfo {author}
  {\bibfnamefont {L.~C.}\ \bibnamefont {Hollenberg}},\ }\href {\doibase
  10.1016/j.physrep.2013.02.001} {\bibfield  {journal} {\bibinfo  {journal}
  {Phys. Rep.}\ }\textbf {\bibinfo {volume} {528}},\ \bibinfo {pages} {1}
  (\bibinfo {year} {2013})}\BibitemShut {NoStop}%
\bibitem [{\citenamefont {Rondin}\ \emph {et~al.}(2014)\citenamefont {Rondin},
  \citenamefont {Tetienne}, \citenamefont {Hingant}, \citenamefont {Roch},
  \citenamefont {Maletinsky},\ and\ \citenamefont {Jacques}}]{Rondin2014}%
  \BibitemOpen
  \bibfield  {author} {\bibinfo {author} {\bibfnamefont {L.}~\bibnamefont
  {Rondin}}, \bibinfo {author} {\bibfnamefont {J.-P.}\ \bibnamefont
  {Tetienne}}, \bibinfo {author} {\bibfnamefont {T.}~\bibnamefont {Hingant}},
  \bibinfo {author} {\bibfnamefont {J.-F.}\ \bibnamefont {Roch}}, \bibinfo
  {author} {\bibfnamefont {P.}~\bibnamefont {Maletinsky}}, \ and\ \bibinfo
  {author} {\bibfnamefont {V.}~\bibnamefont {Jacques}},\ }\href {\doibase
  10.1088/0034-4885/77/5/056503} {\bibfield  {journal} {\bibinfo  {journal}
  {Reports Prog. Phys.}\ }\textbf {\bibinfo {volume} {77}},\ \bibinfo {pages}
  {056503} (\bibinfo {year} {2014})}\BibitemShut {NoStop}%
\bibitem [{\citenamefont {Rondin}\ \emph {et~al.}(2012)\citenamefont {Rondin},
  \citenamefont {Tetienne}, \citenamefont {Spinicelli}, \citenamefont {{Dal
  Savio}}, \citenamefont {Karrai}, \citenamefont {Dantelle}, \citenamefont
  {Thiaville}, \citenamefont {Rohart}, \citenamefont {Roch},\ and\
  \citenamefont {Jacques}}]{Rondin2012}%
  \BibitemOpen
  \bibfield  {author} {\bibinfo {author} {\bibfnamefont {L.}~\bibnamefont
  {Rondin}}, \bibinfo {author} {\bibfnamefont {J.-P.}\ \bibnamefont
  {Tetienne}}, \bibinfo {author} {\bibfnamefont {P.}~\bibnamefont
  {Spinicelli}}, \bibinfo {author} {\bibfnamefont {C.}~\bibnamefont {{Dal
  Savio}}}, \bibinfo {author} {\bibfnamefont {K.}~\bibnamefont {Karrai}},
  \bibinfo {author} {\bibfnamefont {G.}~\bibnamefont {Dantelle}}, \bibinfo
  {author} {\bibfnamefont {A.}~\bibnamefont {Thiaville}}, \bibinfo {author}
  {\bibfnamefont {S.}~\bibnamefont {Rohart}}, \bibinfo {author} {\bibfnamefont
  {J.-F.}\ \bibnamefont {Roch}}, \ and\ \bibinfo {author} {\bibfnamefont
  {V.}~\bibnamefont {Jacques}},\ }\href {\doibase 10.1063/1.3703128} {\bibfield
   {journal} {\bibinfo  {journal} {Appl. Phys. Lett.}\ }\textbf {\bibinfo
  {volume} {100}},\ \bibinfo {pages} {153118} (\bibinfo {year}
  {2012})}\BibitemShut {NoStop}%
\bibitem [{\citenamefont {Maletinsky}\ \emph {et~al.}(2012)\citenamefont
  {Maletinsky}, \citenamefont {Hong}, \citenamefont {Grinolds}, \citenamefont
  {Hausmann}, \citenamefont {Lukin}, \citenamefont {Walsworth}, \citenamefont
  {Loncar},\ and\ \citenamefont {Yacoby}}]{Maletinsky2012}%
  \BibitemOpen
  \bibfield  {author} {\bibinfo {author} {\bibfnamefont {P.}~\bibnamefont
  {Maletinsky}}, \bibinfo {author} {\bibfnamefont {S.}~\bibnamefont {Hong}},
  \bibinfo {author} {\bibfnamefont {M.~S.}\ \bibnamefont {Grinolds}}, \bibinfo
  {author} {\bibfnamefont {B.}~\bibnamefont {Hausmann}}, \bibinfo {author}
  {\bibfnamefont {M.~D.}\ \bibnamefont {Lukin}}, \bibinfo {author}
  {\bibfnamefont {R.~L.}\ \bibnamefont {Walsworth}}, \bibinfo {author}
  {\bibfnamefont {M.}~\bibnamefont {Loncar}}, \ and\ \bibinfo {author}
  {\bibfnamefont {A.}~\bibnamefont {Yacoby}},\ }\href {\doibase
  10.1038/nnano.2012.50} {\bibfield  {journal} {\bibinfo  {journal} {Nat.
  Nanotechnol.}\ }\textbf {\bibinfo {volume} {7}},\ \bibinfo {pages} {320}
  (\bibinfo {year} {2012})}\BibitemShut {NoStop}%
\bibitem [{\citenamefont {Steinert}\ \emph {et~al.}(2010)\citenamefont
  {Steinert}, \citenamefont {Dolde}, \citenamefont {Neumann}, \citenamefont
  {Aird}, \citenamefont {Naydenov}, \citenamefont {Balasubramanian},
  \citenamefont {Jelezko},\ and\ \citenamefont {Wrachtrup}}]{Steinert2010}%
  \BibitemOpen
  \bibfield  {author} {\bibinfo {author} {\bibfnamefont {S.}~\bibnamefont
  {Steinert}}, \bibinfo {author} {\bibfnamefont {F.}~\bibnamefont {Dolde}},
  \bibinfo {author} {\bibfnamefont {P.}~\bibnamefont {Neumann}}, \bibinfo
  {author} {\bibfnamefont {A.}~\bibnamefont {Aird}}, \bibinfo {author}
  {\bibfnamefont {B.}~\bibnamefont {Naydenov}}, \bibinfo {author}
  {\bibfnamefont {G.}~\bibnamefont {Balasubramanian}}, \bibinfo {author}
  {\bibfnamefont {F.}~\bibnamefont {Jelezko}}, \ and\ \bibinfo {author}
  {\bibfnamefont {J.}~\bibnamefont {Wrachtrup}},\ }\href {\doibase
  10.1063/1.3385689} {\bibfield  {journal} {\bibinfo  {journal} {Rev. Sci.
  Instrum.}\ }\textbf {\bibinfo {volume} {81}},\ \bibinfo {pages} {43705}
  (\bibinfo {year} {2010})}\BibitemShut {NoStop}%
\bibitem [{\citenamefont {Pham}\ \emph {et~al.}(2011)\citenamefont {Pham},
  \citenamefont {Sage}, \citenamefont {Stanwix}, \citenamefont {Yeung},
  \citenamefont {Glenn}, \citenamefont {Trifonov}, \citenamefont {Cappellaro},
  \citenamefont {Hemmer}, \citenamefont {Lukin}, \citenamefont {Park},
  \citenamefont {Yacoby},\ and\ \citenamefont {Walsworth}}]{Pham2011}%
  \BibitemOpen
  \bibfield  {author} {\bibinfo {author} {\bibfnamefont {L.~M.}\ \bibnamefont
  {Pham}}, \bibinfo {author} {\bibfnamefont {D.~L.}\ \bibnamefont {Sage}},
  \bibinfo {author} {\bibfnamefont {P.~L.}\ \bibnamefont {Stanwix}}, \bibinfo
  {author} {\bibfnamefont {T.~K.}\ \bibnamefont {Yeung}}, \bibinfo {author}
  {\bibfnamefont {D.}~\bibnamefont {Glenn}}, \bibinfo {author} {\bibfnamefont
  {A.}~\bibnamefont {Trifonov}}, \bibinfo {author} {\bibfnamefont
  {P.}~\bibnamefont {Cappellaro}}, \bibinfo {author} {\bibfnamefont {P.~R.}\
  \bibnamefont {Hemmer}}, \bibinfo {author} {\bibfnamefont {M.~D.}\
  \bibnamefont {Lukin}}, \bibinfo {author} {\bibfnamefont {H.}~\bibnamefont
  {Park}}, \bibinfo {author} {\bibfnamefont {A.}~\bibnamefont {Yacoby}}, \ and\
  \bibinfo {author} {\bibfnamefont {R.~L.}\ \bibnamefont {Walsworth}},\ }\href
  {\doibase 10.1088/1367-2630/13/4/045021} {\bibfield  {journal} {\bibinfo
  {journal} {New J. Phys.}\ }\textbf {\bibinfo {volume} {13}},\ \bibinfo
  {pages} {045021} (\bibinfo {year} {2011})}\BibitemShut {NoStop}%
\bibitem [{\citenamefont {Tetienne}\ \emph {et~al.}(2017)\citenamefont
  {Tetienne}, \citenamefont {Dontschuk}, \citenamefont {Broadway},
  \citenamefont {Stacey}, \citenamefont {Simpson},\ and\ \citenamefont
  {Hollenberg}}]{Tetienne2017a}%
  \BibitemOpen
  \bibfield  {author} {\bibinfo {author} {\bibfnamefont {J.-P.}\ \bibnamefont
  {Tetienne}}, \bibinfo {author} {\bibfnamefont {N.}~\bibnamefont {Dontschuk}},
  \bibinfo {author} {\bibfnamefont {D.~A.}\ \bibnamefont {Broadway}}, \bibinfo
  {author} {\bibfnamefont {A.}~\bibnamefont {Stacey}}, \bibinfo {author}
  {\bibfnamefont {D.~A.}\ \bibnamefont {Simpson}}, \ and\ \bibinfo {author}
  {\bibfnamefont {L.~C.~L.}\ \bibnamefont {Hollenberg}},\ }\href {\doibase
  10.1126/sciadv.1602429} {\bibfield  {journal} {\bibinfo  {journal} {Sci.
  Adv.}\ }\textbf {\bibinfo {volume} {3}},\ \bibinfo {pages} {e1602429}
  (\bibinfo {year} {2017})}\BibitemShut {NoStop}%
\bibitem [{\citenamefont {Uri}\ \emph {et~al.}(2020)\citenamefont {Uri},
  \citenamefont {Kim}, \citenamefont {Bagani}, \citenamefont {Lewandowski},
  \citenamefont {Grover}, \citenamefont {Auerbach}, \citenamefont {Lachman},
  \citenamefont {Myasoedov}, \citenamefont {Taniguchi}, \citenamefont
  {Watanabe}, \citenamefont {Smet},\ and\ \citenamefont {Zeldov}}]{Uri2019}%
  \BibitemOpen
  \bibfield  {author} {\bibinfo {author} {\bibfnamefont {A.}~\bibnamefont
  {Uri}}, \bibinfo {author} {\bibfnamefont {Y.}~\bibnamefont {Kim}}, \bibinfo
  {author} {\bibfnamefont {K.}~\bibnamefont {Bagani}}, \bibinfo {author}
  {\bibfnamefont {C.~K.}\ \bibnamefont {Lewandowski}}, \bibinfo {author}
  {\bibfnamefont {S.}~\bibnamefont {Grover}}, \bibinfo {author} {\bibfnamefont
  {N.}~\bibnamefont {Auerbach}}, \bibinfo {author} {\bibfnamefont {E.~O.}\
  \bibnamefont {Lachman}}, \bibinfo {author} {\bibfnamefont {Y.}~\bibnamefont
  {Myasoedov}}, \bibinfo {author} {\bibfnamefont {T.}~\bibnamefont
  {Taniguchi}}, \bibinfo {author} {\bibfnamefont {K.}~\bibnamefont {Watanabe}},
  \bibinfo {author} {\bibfnamefont {J.}~\bibnamefont {Smet}}, \ and\ \bibinfo
  {author} {\bibfnamefont {E.}~\bibnamefont {Zeldov}},\ }\href {\doibase
  10.1038/s41567-019-0713-3} {\bibfield  {journal} {\bibinfo  {journal} {Nat.
  Phys.}\ }\textbf {\bibinfo {volume} {16}},\ \bibinfo {pages} {164} (\bibinfo
  {year} {2020})}\BibitemShut {NoStop}%
\bibitem [{\citenamefont {Du}\ \emph {et~al.}(2017)\citenamefont {Du},
  \citenamefont {van~der Sar}, \citenamefont {Zhou}, \citenamefont {Upadhyaya},
  \citenamefont {Casola}, \citenamefont {Zhang}, \citenamefont {Onbasli},
  \citenamefont {Ross}, \citenamefont {Walsworth}, \citenamefont
  {Tserkovnyak},\ and\ \citenamefont {Yacoby}}]{Du2017}%
  \BibitemOpen
  \bibfield  {author} {\bibinfo {author} {\bibfnamefont {C.}~\bibnamefont
  {Du}}, \bibinfo {author} {\bibfnamefont {T.}~\bibnamefont {van~der Sar}},
  \bibinfo {author} {\bibfnamefont {T.~X.}\ \bibnamefont {Zhou}}, \bibinfo
  {author} {\bibfnamefont {P.}~\bibnamefont {Upadhyaya}}, \bibinfo {author}
  {\bibfnamefont {F.}~\bibnamefont {Casola}}, \bibinfo {author} {\bibfnamefont
  {H.}~\bibnamefont {Zhang}}, \bibinfo {author} {\bibfnamefont {M.~C.}\
  \bibnamefont {Onbasli}}, \bibinfo {author} {\bibfnamefont {C.~A.}\
  \bibnamefont {Ross}}, \bibinfo {author} {\bibfnamefont {R.~L.}\ \bibnamefont
  {Walsworth}}, \bibinfo {author} {\bibfnamefont {Y.}~\bibnamefont
  {Tserkovnyak}}, \ and\ \bibinfo {author} {\bibfnamefont {A.}~\bibnamefont
  {Yacoby}},\ }\href {\doibase 10.1126/science.aak9611} {\bibfield  {journal}
  {\bibinfo  {journal} {Science}\ }\textbf {\bibinfo {volume} {357}},\ \bibinfo
  {pages} {195} (\bibinfo {year} {2017})}\BibitemShut {NoStop}%
\bibitem [{\citenamefont {Anahory}\ \emph {et~al.}(2016)\citenamefont
  {Anahory}, \citenamefont {Embon}, \citenamefont {Li}, \citenamefont
  {Banerjee}, \citenamefont {Meltzer}, \citenamefont {Naren}, \citenamefont
  {Yakovenko}, \citenamefont {Cuppens}, \citenamefont {Myasoedov},
  \citenamefont {Rappaport}, \citenamefont {Huber}, \citenamefont {Michaeli},
  \citenamefont {Venkatesan}, \citenamefont {Ariando},\ and\ \citenamefont
  {Zeldov}}]{Anahory2016}%
  \BibitemOpen
  \bibfield  {author} {\bibinfo {author} {\bibfnamefont {Y.}~\bibnamefont
  {Anahory}}, \bibinfo {author} {\bibfnamefont {L.}~\bibnamefont {Embon}},
  \bibinfo {author} {\bibfnamefont {C.~J.}\ \bibnamefont {Li}}, \bibinfo
  {author} {\bibfnamefont {S.}~\bibnamefont {Banerjee}}, \bibinfo {author}
  {\bibfnamefont {A.}~\bibnamefont {Meltzer}}, \bibinfo {author} {\bibfnamefont
  {H.~R.}\ \bibnamefont {Naren}}, \bibinfo {author} {\bibfnamefont
  {A.}~\bibnamefont {Yakovenko}}, \bibinfo {author} {\bibfnamefont
  {J.}~\bibnamefont {Cuppens}}, \bibinfo {author} {\bibfnamefont
  {Y.}~\bibnamefont {Myasoedov}}, \bibinfo {author} {\bibfnamefont {M.~L.}\
  \bibnamefont {Rappaport}}, \bibinfo {author} {\bibfnamefont {M.~E.}\
  \bibnamefont {Huber}}, \bibinfo {author} {\bibfnamefont {K.}~\bibnamefont
  {Michaeli}}, \bibinfo {author} {\bibfnamefont {T.}~\bibnamefont
  {Venkatesan}}, \bibinfo {author} {\bibnamefont {Ariando}}, \ and\ \bibinfo
  {author} {\bibfnamefont {E.}~\bibnamefont {Zeldov}},\ }\href {\doibase
  10.1038/ncomms12566} {\bibfield  {journal} {\bibinfo  {journal} {Nat.
  Commun.}\ }\textbf {\bibinfo {volume} {7}},\ \bibinfo {pages} {12566}
  (\bibinfo {year} {2016})}\BibitemShut {NoStop}%
\bibitem [{\citenamefont {Gibertini}\ \emph {et~al.}(2019)\citenamefont
  {Gibertini}, \citenamefont {Koperski}, \citenamefont {Morpurgo},\ and\
  \citenamefont {Novoselov}}]{Gibertini2019}%
  \BibitemOpen
  \bibfield  {author} {\bibinfo {author} {\bibfnamefont {M.}~\bibnamefont
  {Gibertini}}, \bibinfo {author} {\bibfnamefont {M.}~\bibnamefont {Koperski}},
  \bibinfo {author} {\bibfnamefont {A.~F.}\ \bibnamefont {Morpurgo}}, \ and\
  \bibinfo {author} {\bibfnamefont {K.~S.}\ \bibnamefont {Novoselov}},\ }\href
  {\doibase 10.1038/s41565-019-0438-6} {\bibfield  {journal} {\bibinfo
  {journal} {Nat. Nanotechnol.}\ }\textbf {\bibinfo {volume} {14}},\ \bibinfo
  {pages} {408} (\bibinfo {year} {2019})}\BibitemShut {NoStop}%
\bibitem [{\citenamefont {Gong}\ and\ \citenamefont {Zhang}(2019)}]{Gong2019}%
  \BibitemOpen
  \bibfield  {author} {\bibinfo {author} {\bibfnamefont {C.}~\bibnamefont
  {Gong}}\ and\ \bibinfo {author} {\bibfnamefont {X.}~\bibnamefont {Zhang}},\
  }\href {\doibase 10.1126/science.aav4450} {\bibfield  {journal} {\bibinfo
  {journal} {Science}\ }\textbf {\bibinfo {volume} {363}},\ \bibinfo {pages}
  {706} (\bibinfo {year} {2019})}\BibitemShut {NoStop}%
\bibitem [{\citenamefont {Thiel}\ \emph {et~al.}(2019)\citenamefont {Thiel},
  \citenamefont {Wang}, \citenamefont {Tschudin}, \citenamefont {Rohner},
  \citenamefont {Guti{\'{e}}rrez-Lezama}, \citenamefont {Ubrig}, \citenamefont
  {Gibertini}, \citenamefont {Giannini}, \citenamefont {Morpurgo},\ and\
  \citenamefont {Maletinsky}}]{Thiel2019}%
  \BibitemOpen
  \bibfield  {author} {\bibinfo {author} {\bibfnamefont {L.}~\bibnamefont
  {Thiel}}, \bibinfo {author} {\bibfnamefont {Z.}~\bibnamefont {Wang}},
  \bibinfo {author} {\bibfnamefont {M.~A.}\ \bibnamefont {Tschudin}}, \bibinfo
  {author} {\bibfnamefont {D.}~\bibnamefont {Rohner}}, \bibinfo {author}
  {\bibfnamefont {I.}~\bibnamefont {Guti{\'{e}}rrez-Lezama}}, \bibinfo {author}
  {\bibfnamefont {N.}~\bibnamefont {Ubrig}}, \bibinfo {author} {\bibfnamefont
  {M.}~\bibnamefont {Gibertini}}, \bibinfo {author} {\bibfnamefont
  {E.}~\bibnamefont {Giannini}}, \bibinfo {author} {\bibfnamefont {A.~F.}\
  \bibnamefont {Morpurgo}}, \ and\ \bibinfo {author} {\bibfnamefont
  {P.}~\bibnamefont {Maletinsky}},\ }\href {\doibase 10.1126/science.aav6926}
  {\bibfield  {journal} {\bibinfo  {journal} {Science}\ }\textbf {\bibinfo
  {volume} {364}},\ \bibinfo {pages} {973} (\bibinfo {year}
  {2019})}\BibitemShut {NoStop}%
\bibitem [{\citenamefont {Broadway}\ \emph {et~al.}()\citenamefont {Broadway},
  \citenamefont {Scholten}, \citenamefont {Tan}, \citenamefont {Dontschuk},
  \citenamefont {Lillie}, \citenamefont {Johnson}, \citenamefont {Zheng},
  \citenamefont {Wang}, \citenamefont {Oganov}, \citenamefont {Tian},
  \citenamefont {Li}, \citenamefont {Lei}, \citenamefont {Wang}, \citenamefont
  {Hollenberg},\ and\ \citenamefont {Tetienne}}]{Broadway2020}%
  \BibitemOpen
  \bibfield  {author} {\bibinfo {author} {\bibfnamefont {D.~A.}\ \bibnamefont
  {Broadway}}, \bibinfo {author} {\bibfnamefont {S.~C.}\ \bibnamefont
  {Scholten}}, \bibinfo {author} {\bibfnamefont {C.}~\bibnamefont {Tan}},
  \bibinfo {author} {\bibfnamefont {N.}~\bibnamefont {Dontschuk}}, \bibinfo
  {author} {\bibfnamefont {S.~E.}\ \bibnamefont {Lillie}}, \bibinfo {author}
  {\bibfnamefont {B.~C.}\ \bibnamefont {Johnson}}, \bibinfo {author}
  {\bibfnamefont {G.}~\bibnamefont {Zheng}}, \bibinfo {author} {\bibfnamefont
  {Z.}~\bibnamefont {Wang}}, \bibinfo {author} {\bibfnamefont {A.~R.}\
  \bibnamefont {Oganov}}, \bibinfo {author} {\bibfnamefont {S.}~\bibnamefont
  {Tian}}, \bibinfo {author} {\bibfnamefont {C.}~\bibnamefont {Li}}, \bibinfo
  {author} {\bibfnamefont {H.}~\bibnamefont {Lei}}, \bibinfo {author}
  {\bibfnamefont {L.}~\bibnamefont {Wang}}, \bibinfo {author} {\bibfnamefont
  {L.~C.~L.}\ \bibnamefont {Hollenberg}}, \ and\ \bibinfo {author}
  {\bibfnamefont {J.-P.}\ \bibnamefont {Tetienne}},\ }\href
  {http://arxiv.org/abs/2003.08470} {\ }\Eprint
  {http://arxiv.org/abs/2003.08470} {arXiv:2003.08470} \BibitemShut {NoStop}%
\bibitem [{\citenamefont {W{\"{o}}rnle}\ \emph {et~al.}()\citenamefont
  {W{\"{o}}rnle}, \citenamefont {Welter}, \citenamefont {Ka{\v{s}}par},
  \citenamefont {Olejn{\'{i}}k}, \citenamefont {Nov{\'{a}}k}, \citenamefont
  {Campion}, \citenamefont {Wadley}, \citenamefont {Jungwirth}, \citenamefont
  {Degen},\ and\ \citenamefont {Gambardella}}]{Wornle2019}%
  \BibitemOpen
  \bibfield  {author} {\bibinfo {author} {\bibfnamefont {M.~S.}\ \bibnamefont
  {W{\"{o}}rnle}}, \bibinfo {author} {\bibfnamefont {P.}~\bibnamefont
  {Welter}}, \bibinfo {author} {\bibfnamefont {Z.}~\bibnamefont
  {Ka{\v{s}}par}}, \bibinfo {author} {\bibfnamefont {K.}~\bibnamefont
  {Olejn{\'{i}}k}}, \bibinfo {author} {\bibfnamefont {V.}~\bibnamefont
  {Nov{\'{a}}k}}, \bibinfo {author} {\bibfnamefont {R.~P.}\ \bibnamefont
  {Campion}}, \bibinfo {author} {\bibfnamefont {P.}~\bibnamefont {Wadley}},
  \bibinfo {author} {\bibfnamefont {T.}~\bibnamefont {Jungwirth}}, \bibinfo
  {author} {\bibfnamefont {C.~L.}\ \bibnamefont {Degen}}, \ and\ \bibinfo
  {author} {\bibfnamefont {P.}~\bibnamefont {Gambardella}},\ }\href
  {http://arxiv.org/abs/1912.05287} {\ }\Eprint
  {http://arxiv.org/abs/1912.05287} {arXiv:1912.05287} \BibitemShut {NoStop}%
\bibitem [{\citenamefont {Kim}\ \emph {et~al.}(2019)\citenamefont {Kim},
  \citenamefont {Kumaravadivel}, \citenamefont {Birkbeck}, \citenamefont
  {Kuang}, \citenamefont {Xu}, \citenamefont {Hopkinson}, \citenamefont
  {Knolle}, \citenamefont {McClarty}, \citenamefont {Berdyugin}, \citenamefont
  {{Ben Shalom}}, \citenamefont {Gorbachev}, \citenamefont {Haigh},
  \citenamefont {Liu}, \citenamefont {Edgar}, \citenamefont {Novoselov},
  \citenamefont {Grigorieva},\ and\ \citenamefont {Geim}}]{Kim2019}%
  \BibitemOpen
  \bibfield  {author} {\bibinfo {author} {\bibfnamefont {M.}~\bibnamefont
  {Kim}}, \bibinfo {author} {\bibfnamefont {P.}~\bibnamefont {Kumaravadivel}},
  \bibinfo {author} {\bibfnamefont {J.}~\bibnamefont {Birkbeck}}, \bibinfo
  {author} {\bibfnamefont {W.}~\bibnamefont {Kuang}}, \bibinfo {author}
  {\bibfnamefont {S.~G.}\ \bibnamefont {Xu}}, \bibinfo {author} {\bibfnamefont
  {D.~G.}\ \bibnamefont {Hopkinson}}, \bibinfo {author} {\bibfnamefont
  {J.}~\bibnamefont {Knolle}}, \bibinfo {author} {\bibfnamefont {P.~A.}\
  \bibnamefont {McClarty}}, \bibinfo {author} {\bibfnamefont {A.~I.}\
  \bibnamefont {Berdyugin}}, \bibinfo {author} {\bibfnamefont {M.}~\bibnamefont
  {{Ben Shalom}}}, \bibinfo {author} {\bibfnamefont {R.~V.}\ \bibnamefont
  {Gorbachev}}, \bibinfo {author} {\bibfnamefont {S.~J.}\ \bibnamefont
  {Haigh}}, \bibinfo {author} {\bibfnamefont {S.}~\bibnamefont {Liu}}, \bibinfo
  {author} {\bibfnamefont {J.~H.}\ \bibnamefont {Edgar}}, \bibinfo {author}
  {\bibfnamefont {K.~S.}\ \bibnamefont {Novoselov}}, \bibinfo {author}
  {\bibfnamefont {I.~V.}\ \bibnamefont {Grigorieva}}, \ and\ \bibinfo {author}
  {\bibfnamefont {A.~K.}\ \bibnamefont {Geim}},\ }\href {\doibase
  10.1038/s41928-019-0302-6} {\bibfield  {journal} {\bibinfo  {journal} {Nat.
  Electron.}\ }\textbf {\bibinfo {volume} {2}},\ \bibinfo {pages} {457}
  (\bibinfo {year} {2019})}\BibitemShut {NoStop}%
\bibitem [{\citenamefont {Gross}\ \emph {et~al.}(2017)\citenamefont {Gross},
  \citenamefont {Akhtar}, \citenamefont {Garcia}, \citenamefont
  {Mart{\'{i}}nez}, \citenamefont {Chouaieb}, \citenamefont {Garcia},
  \citenamefont {Carr{\'{e}}t{\'{e}}ro}, \citenamefont
  {Barth{\'{e}}l{\'{e}}my}, \citenamefont {Appel}, \citenamefont {Maletinsky},
  \citenamefont {Kim}, \citenamefont {Chauleau}, \citenamefont {Jaouen},
  \citenamefont {Viret}, \citenamefont {Bibes}, \citenamefont {Fusil},\ and\
  \citenamefont {Jacques}}]{Gross2017a}%
  \BibitemOpen
  \bibfield  {author} {\bibinfo {author} {\bibfnamefont {I.}~\bibnamefont
  {Gross}}, \bibinfo {author} {\bibfnamefont {W.}~\bibnamefont {Akhtar}},
  \bibinfo {author} {\bibfnamefont {V.}~\bibnamefont {Garcia}}, \bibinfo
  {author} {\bibfnamefont {L.~J.}\ \bibnamefont {Mart{\'{i}}nez}}, \bibinfo
  {author} {\bibfnamefont {S.}~\bibnamefont {Chouaieb}}, \bibinfo {author}
  {\bibfnamefont {K.}~\bibnamefont {Garcia}}, \bibinfo {author} {\bibfnamefont
  {C.}~\bibnamefont {Carr{\'{e}}t{\'{e}}ro}}, \bibinfo {author} {\bibfnamefont
  {A.}~\bibnamefont {Barth{\'{e}}l{\'{e}}my}}, \bibinfo {author} {\bibfnamefont
  {P.}~\bibnamefont {Appel}}, \bibinfo {author} {\bibfnamefont
  {P.}~\bibnamefont {Maletinsky}}, \bibinfo {author} {\bibfnamefont {J.-V.}\
  \bibnamefont {Kim}}, \bibinfo {author} {\bibfnamefont {J.~Y.}\ \bibnamefont
  {Chauleau}}, \bibinfo {author} {\bibfnamefont {N.}~\bibnamefont {Jaouen}},
  \bibinfo {author} {\bibfnamefont {M.}~\bibnamefont {Viret}}, \bibinfo
  {author} {\bibfnamefont {M.}~\bibnamefont {Bibes}}, \bibinfo {author}
  {\bibfnamefont {S.}~\bibnamefont {Fusil}}, \ and\ \bibinfo {author}
  {\bibfnamefont {V.}~\bibnamefont {Jacques}},\ }\href {\doibase
  10.1038/nature23656} {\bibfield  {journal} {\bibinfo  {journal} {Nature}\
  }\textbf {\bibinfo {volume} {549}},\ \bibinfo {pages} {252} (\bibinfo {year}
  {2017})}\BibitemShut {NoStop}%
\bibitem [{\citenamefont {Tetienne}\ \emph {et~al.}(2014)\citenamefont
  {Tetienne}, \citenamefont {Hingant}, \citenamefont {Kim}, \citenamefont
  {Diez}, \citenamefont {Adam}, \citenamefont {Garcia}, \citenamefont {Roch},
  \citenamefont {Rohart}, \citenamefont {Thiaville}, \citenamefont
  {Ravelosona},\ and\ \citenamefont {Jacques}}]{Tetienne2014}%
  \BibitemOpen
  \bibfield  {author} {\bibinfo {author} {\bibfnamefont {J.~P.}\ \bibnamefont
  {Tetienne}}, \bibinfo {author} {\bibfnamefont {T.}~\bibnamefont {Hingant}},
  \bibinfo {author} {\bibfnamefont {J.~V.}\ \bibnamefont {Kim}}, \bibinfo
  {author} {\bibfnamefont {L.~H.}\ \bibnamefont {Diez}}, \bibinfo {author}
  {\bibfnamefont {J.~P.}\ \bibnamefont {Adam}}, \bibinfo {author}
  {\bibfnamefont {K.}~\bibnamefont {Garcia}}, \bibinfo {author} {\bibfnamefont
  {J.~F.}\ \bibnamefont {Roch}}, \bibinfo {author} {\bibfnamefont
  {S.}~\bibnamefont {Rohart}}, \bibinfo {author} {\bibfnamefont
  {A.}~\bibnamefont {Thiaville}}, \bibinfo {author} {\bibfnamefont
  {D.}~\bibnamefont {Ravelosona}}, \ and\ \bibinfo {author} {\bibfnamefont
  {V.}~\bibnamefont {Jacques}},\ }\href {\doibase 10.1126/science.1250113}
  {\bibfield  {journal} {\bibinfo  {journal} {Science}\ }\textbf {\bibinfo
  {volume} {344}},\ \bibinfo {pages} {1366} (\bibinfo {year}
  {2014})}\BibitemShut {NoStop}%
\bibitem [{\citenamefont {Tetienne}\ \emph {et~al.}(2015)\citenamefont
  {Tetienne}, \citenamefont {Hingant}, \citenamefont {Mart{\'{i}}nez},
  \citenamefont {Rohart}, \citenamefont {Thiaville}, \citenamefont {Diez},
  \citenamefont {Garcia}, \citenamefont {Adam}, \citenamefont {Kim},
  \citenamefont {Roch}, \citenamefont {Miron}, \citenamefont {Gaudin},
  \citenamefont {Vila}, \citenamefont {Ocker}, \citenamefont {Ravelosona},\
  and\ \citenamefont {Jacques}}]{Tetienne2015}%
  \BibitemOpen
  \bibfield  {author} {\bibinfo {author} {\bibfnamefont {J.-P.}\ \bibnamefont
  {Tetienne}}, \bibinfo {author} {\bibfnamefont {T.}~\bibnamefont {Hingant}},
  \bibinfo {author} {\bibfnamefont {L.}~\bibnamefont {Mart{\'{i}}nez}},
  \bibinfo {author} {\bibfnamefont {S.}~\bibnamefont {Rohart}}, \bibinfo
  {author} {\bibfnamefont {A.}~\bibnamefont {Thiaville}}, \bibinfo {author}
  {\bibfnamefont {L.~H.}\ \bibnamefont {Diez}}, \bibinfo {author}
  {\bibfnamefont {K.}~\bibnamefont {Garcia}}, \bibinfo {author} {\bibfnamefont
  {J.-P.}\ \bibnamefont {Adam}}, \bibinfo {author} {\bibfnamefont {J.-V.}\
  \bibnamefont {Kim}}, \bibinfo {author} {\bibfnamefont {J.-F.}\ \bibnamefont
  {Roch}}, \bibinfo {author} {\bibfnamefont {I.}~\bibnamefont {Miron}},
  \bibinfo {author} {\bibfnamefont {G.}~\bibnamefont {Gaudin}}, \bibinfo
  {author} {\bibfnamefont {L.}~\bibnamefont {Vila}}, \bibinfo {author}
  {\bibfnamefont {B.}~\bibnamefont {Ocker}}, \bibinfo {author} {\bibfnamefont
  {D.}~\bibnamefont {Ravelosona}}, \ and\ \bibinfo {author} {\bibfnamefont
  {V.}~\bibnamefont {Jacques}},\ }\href {\doibase 10.1038/ncomms7733}
  {\bibfield  {journal} {\bibinfo  {journal} {Nat. Commun.}\ }\textbf {\bibinfo
  {volume} {6}},\ \bibinfo {pages} {6733} (\bibinfo {year} {2015})}\BibitemShut
  {NoStop}%
\bibitem [{\citenamefont {Bandurin}\ \emph {et~al.}(2016)\citenamefont
  {Bandurin}, \citenamefont {Torre}, \citenamefont {Kumar}, \citenamefont {{Ben
  Shalom}}, \citenamefont {Tomadin}, \citenamefont {Principi}, \citenamefont
  {Auton}, \citenamefont {Khestanova}, \citenamefont {Novoselov}, \citenamefont
  {Grigorieva}, \citenamefont {Ponomarenko}, \citenamefont {Geim},\ and\
  \citenamefont {Polini}}]{Bandurin2016}%
  \BibitemOpen
  \bibfield  {author} {\bibinfo {author} {\bibfnamefont {D.~A.}\ \bibnamefont
  {Bandurin}}, \bibinfo {author} {\bibfnamefont {I.}~\bibnamefont {Torre}},
  \bibinfo {author} {\bibfnamefont {R.~K.}\ \bibnamefont {Kumar}}, \bibinfo
  {author} {\bibfnamefont {M.}~\bibnamefont {{Ben Shalom}}}, \bibinfo {author}
  {\bibfnamefont {A.}~\bibnamefont {Tomadin}}, \bibinfo {author} {\bibfnamefont
  {A.}~\bibnamefont {Principi}}, \bibinfo {author} {\bibfnamefont {G.~H.}\
  \bibnamefont {Auton}}, \bibinfo {author} {\bibfnamefont {E.}~\bibnamefont
  {Khestanova}}, \bibinfo {author} {\bibfnamefont {K.~S.}\ \bibnamefont
  {Novoselov}}, \bibinfo {author} {\bibfnamefont {I.~V.}\ \bibnamefont
  {Grigorieva}}, \bibinfo {author} {\bibfnamefont {L.~A.}\ \bibnamefont
  {Ponomarenko}}, \bibinfo {author} {\bibfnamefont {A.~K.}\ \bibnamefont
  {Geim}}, \ and\ \bibinfo {author} {\bibfnamefont {M.}~\bibnamefont
  {Polini}},\ }\href {\doibase 10.1126/science.aad0201} {\bibfield  {journal}
  {\bibinfo  {journal} {Science}\ }\textbf {\bibinfo {volume} {351}},\ \bibinfo
  {pages} {1055} (\bibinfo {year} {2016})}\BibitemShut {NoStop}%
\bibitem [{\citenamefont {Cheianov}\ \emph {et~al.}(2007)\citenamefont
  {Cheianov}, \citenamefont {Fal'ko},\ and\ \citenamefont
  {Altshuler}}]{Cheianov2007}%
  \BibitemOpen
  \bibfield  {author} {\bibinfo {author} {\bibfnamefont {V.~V.}\ \bibnamefont
  {Cheianov}}, \bibinfo {author} {\bibfnamefont {V.}~\bibnamefont {Fal'ko}}, \
  and\ \bibinfo {author} {\bibfnamefont {B.~L.}\ \bibnamefont {Altshuler}},\
  }\href {\doibase 10.1126/science.1138020} {\bibfield  {journal} {\bibinfo
  {journal} {Science}\ }\textbf {\bibinfo {volume} {315}},\ \bibinfo {pages}
  {1252} (\bibinfo {year} {2007})}\BibitemShut {NoStop}%
\bibitem [{\citenamefont {Chen}\ \emph {et~al.}(2016)\citenamefont {Chen},
  \citenamefont {Han}, \citenamefont {Elahi}, \citenamefont {Habib},
  \citenamefont {Wang}, \citenamefont {Wen}, \citenamefont {Gao}, \citenamefont
  {Taniguchi}, \citenamefont {Watanabe}, \citenamefont {Hone}, \citenamefont
  {Ghosh},\ and\ \citenamefont {Dean}}]{Chen2016b}%
  \BibitemOpen
  \bibfield  {author} {\bibinfo {author} {\bibfnamefont {S.}~\bibnamefont
  {Chen}}, \bibinfo {author} {\bibfnamefont {Z.}~\bibnamefont {Han}}, \bibinfo
  {author} {\bibfnamefont {M.~M.}\ \bibnamefont {Elahi}}, \bibinfo {author}
  {\bibfnamefont {K.~M.}\ \bibnamefont {Habib}}, \bibinfo {author}
  {\bibfnamefont {L.}~\bibnamefont {Wang}}, \bibinfo {author} {\bibfnamefont
  {B.}~\bibnamefont {Wen}}, \bibinfo {author} {\bibfnamefont {Y.}~\bibnamefont
  {Gao}}, \bibinfo {author} {\bibfnamefont {T.}~\bibnamefont {Taniguchi}},
  \bibinfo {author} {\bibfnamefont {K.}~\bibnamefont {Watanabe}}, \bibinfo
  {author} {\bibfnamefont {J.}~\bibnamefont {Hone}}, \bibinfo {author}
  {\bibfnamefont {A.~W.}\ \bibnamefont {Ghosh}}, \ and\ \bibinfo {author}
  {\bibfnamefont {C.~R.}\ \bibnamefont {Dean}},\ }\href {\doibase
  10.1126/science.aaf5481} {\bibfield  {journal} {\bibinfo  {journal}
  {Science}\ }\textbf {\bibinfo {volume} {353}},\ \bibinfo {pages} {1522}
  (\bibinfo {year} {2016})}\BibitemShut {NoStop}%
\bibitem [{\citenamefont {Jungwirth}\ \emph {et~al.}(2016)\citenamefont
  {Jungwirth}, \citenamefont {Marti}, \citenamefont {Wadley},\ and\
  \citenamefont {Wunderlich}}]{Jungwirth2016}%
  \BibitemOpen
  \bibfield  {author} {\bibinfo {author} {\bibfnamefont {T.}~\bibnamefont
  {Jungwirth}}, \bibinfo {author} {\bibfnamefont {X.}~\bibnamefont {Marti}},
  \bibinfo {author} {\bibfnamefont {P.}~\bibnamefont {Wadley}}, \ and\ \bibinfo
  {author} {\bibfnamefont {J.}~\bibnamefont {Wunderlich}},\ }\href {\doibase
  10.1038/nnano.2016.18} {\bibfield  {journal} {\bibinfo  {journal} {Nat.
  Nanotechnol.}\ }\textbf {\bibinfo {volume} {11}},\ \bibinfo {pages} {231}
  (\bibinfo {year} {2016})}\BibitemShut {NoStop}%
\bibitem [{\citenamefont {Gorbachev}\ \emph {et~al.}(2014)\citenamefont
  {Gorbachev}, \citenamefont {Song}, \citenamefont {Yu}, \citenamefont
  {Kretinin}, \citenamefont {Withers}, \citenamefont {Cao}, \citenamefont
  {Mishchenko}, \citenamefont {Grigorieva}, \citenamefont {Novoselov},
  \citenamefont {Levitov},\ and\ \citenamefont {Geim}}]{Gorbachev2014}%
  \BibitemOpen
  \bibfield  {author} {\bibinfo {author} {\bibfnamefont {R.~V.}\ \bibnamefont
  {Gorbachev}}, \bibinfo {author} {\bibfnamefont {J.~C.~W.}\ \bibnamefont
  {Song}}, \bibinfo {author} {\bibfnamefont {G.~L.}\ \bibnamefont {Yu}},
  \bibinfo {author} {\bibfnamefont {A.~V.}\ \bibnamefont {Kretinin}}, \bibinfo
  {author} {\bibfnamefont {F.}~\bibnamefont {Withers}}, \bibinfo {author}
  {\bibfnamefont {Y.}~\bibnamefont {Cao}}, \bibinfo {author} {\bibfnamefont
  {A.}~\bibnamefont {Mishchenko}}, \bibinfo {author} {\bibfnamefont {I.~V.}\
  \bibnamefont {Grigorieva}}, \bibinfo {author} {\bibfnamefont {K.~S.}\
  \bibnamefont {Novoselov}}, \bibinfo {author} {\bibfnamefont {L.~S.}\
  \bibnamefont {Levitov}}, \ and\ \bibinfo {author} {\bibfnamefont {A.~K.}\
  \bibnamefont {Geim}},\ }\href {\doibase 10.1126/science.1254966} {\bibfield
  {journal} {\bibinfo  {journal} {Science}\ }\textbf {\bibinfo {volume}
  {346}},\ \bibinfo {pages} {448} (\bibinfo {year} {2014})}\BibitemShut
  {NoStop}%
\bibitem [{\citenamefont {Roth}\ \emph {et~al.}(1989)\citenamefont {Roth},
  \citenamefont {Sepulveda},\ and\ \citenamefont {Wikswo}}]{Roth1989}%
  \BibitemOpen
  \bibfield  {author} {\bibinfo {author} {\bibfnamefont {B.~J.}\ \bibnamefont
  {Roth}}, \bibinfo {author} {\bibfnamefont {N.~G.}\ \bibnamefont {Sepulveda}},
  \ and\ \bibinfo {author} {\bibfnamefont {J.~P.}\ \bibnamefont {Wikswo}},\
  }\href {\doibase 10.1063/1.342549} {\bibfield  {journal} {\bibinfo  {journal}
  {J. Appl. Phys.}\ }\textbf {\bibinfo {volume} {65}},\ \bibinfo {pages} {361}
  (\bibinfo {year} {1989})}\BibitemShut {NoStop}%
\bibitem [{\citenamefont {Meltzer}\ \emph {et~al.}(2017)\citenamefont
  {Meltzer}, \citenamefont {Levin},\ and\ \citenamefont
  {Zeldov}}]{Meltzer2017}%
  \BibitemOpen
  \bibfield  {author} {\bibinfo {author} {\bibfnamefont {A.~Y.}\ \bibnamefont
  {Meltzer}}, \bibinfo {author} {\bibfnamefont {E.}~\bibnamefont {Levin}}, \
  and\ \bibinfo {author} {\bibfnamefont {E.}~\bibnamefont {Zeldov}},\ }\href
  {\doibase 10.1103/PhysRevApplied.8.064030} {\bibfield  {journal} {\bibinfo
  {journal} {Phys. Rev. Appl.}\ }\textbf {\bibinfo {volume} {8}},\ \bibinfo
  {pages} {064030} (\bibinfo {year} {2017})}\BibitemShut {NoStop}%
\bibitem [{\citenamefont {Knauss}\ \emph {et~al.}(2001)\citenamefont {Knauss},
  \citenamefont {Cawthorne}, \citenamefont {Lettsome}, \citenamefont {Kelly},
  \citenamefont {Chatraphorn}, \citenamefont {Fleet}, \citenamefont
  {Wellstood},\ and\ \citenamefont {Vanderlinde}}]{Knauss2001}%
  \BibitemOpen
  \bibfield  {author} {\bibinfo {author} {\bibfnamefont {L.~A.}\ \bibnamefont
  {Knauss}}, \bibinfo {author} {\bibfnamefont {A.~B.}\ \bibnamefont
  {Cawthorne}}, \bibinfo {author} {\bibfnamefont {N.}~\bibnamefont {Lettsome}},
  \bibinfo {author} {\bibfnamefont {S.}~\bibnamefont {Kelly}}, \bibinfo
  {author} {\bibfnamefont {S.}~\bibnamefont {Chatraphorn}}, \bibinfo {author}
  {\bibfnamefont {E.~F.}\ \bibnamefont {Fleet}}, \bibinfo {author}
  {\bibfnamefont {F.~C.}\ \bibnamefont {Wellstood}}, \ and\ \bibinfo {author}
  {\bibfnamefont {W.~E.}\ \bibnamefont {Vanderlinde}},\ }\href {\doibase
  10.1016/S0026-2714(01)00108-1} {\bibfield  {journal} {\bibinfo  {journal}
  {Microelectron. Reliab.}\ }\textbf {\bibinfo {volume} {41}},\ \bibinfo
  {pages} {1211} (\bibinfo {year} {2001})}\BibitemShut {NoStop}%
\bibitem [{\citenamefont {Clement}\ \emph {et~al.}()\citenamefont {Clement},
  \citenamefont {Sethna},\ and\ \citenamefont {Nowack}}]{Clement2019}%
  \BibitemOpen
  \bibfield  {author} {\bibinfo {author} {\bibfnamefont {C.~B.}\ \bibnamefont
  {Clement}}, \bibinfo {author} {\bibfnamefont {J.~P.}\ \bibnamefont {Sethna}},
  \ and\ \bibinfo {author} {\bibfnamefont {K.~C.}\ \bibnamefont {Nowack}},\
  }\href {http://arxiv.org/abs/1910.12929} {\ }\Eprint
  {http://arxiv.org/abs/1910.12929} {arXiv:1910.12929} \BibitemShut {NoStop}%
\bibitem [{\citenamefont {Lima}\ and\ \citenamefont {Weiss}(2009)}]{Lima2009}%
  \BibitemOpen
  \bibfield  {author} {\bibinfo {author} {\bibfnamefont {E.~A.}\ \bibnamefont
  {Lima}}\ and\ \bibinfo {author} {\bibfnamefont {B.~P.}\ \bibnamefont
  {Weiss}},\ }\href {\doibase 10.1029/2008JB006006} {\bibfield  {journal}
  {\bibinfo  {journal} {J. Geophys. Res.}\ }\textbf {\bibinfo {volume} {114}},\
  \bibinfo {pages} {B06102} (\bibinfo {year} {2009})}\BibitemShut {NoStop}%
\bibitem [{\citenamefont {Casola}\ \emph {et~al.}(2018)\citenamefont {Casola},
  \citenamefont {van~der Sar},\ and\ \citenamefont {Yacoby}}]{Casola2018}%
  \BibitemOpen
  \bibfield  {author} {\bibinfo {author} {\bibfnamefont {F.}~\bibnamefont
  {Casola}}, \bibinfo {author} {\bibfnamefont {T.}~\bibnamefont {van~der Sar}},
  \ and\ \bibinfo {author} {\bibfnamefont {A.}~\bibnamefont {Yacoby}},\ }\href
  {\doibase 10.1038/natrevmats.2017.88} {\bibfield  {journal} {\bibinfo
  {journal} {Nat. Rev. Mater.}\ }\textbf {\bibinfo {volume} {3}},\ \bibinfo
  {pages} {17088} (\bibinfo {year} {2018})}\BibitemShut {NoStop}%
\bibitem [{\citenamefont {{Shaofen Tan}}\ \emph {et~al.}(1996)\citenamefont
  {{Shaofen Tan}}, \citenamefont {{Yu Pei Ma}}, \citenamefont {Thomas},\ and\
  \citenamefont {Wikswo}}]{Tan1996}%
  \BibitemOpen
  \bibfield  {author} {\bibinfo {author} {\bibnamefont {{Shaofen Tan}}},
  \bibinfo {author} {\bibnamefont {{Yu Pei Ma}}}, \bibinfo {author}
  {\bibfnamefont {I.}~\bibnamefont {Thomas}}, \ and\ \bibinfo {author}
  {\bibfnamefont {J.}~\bibnamefont {Wikswo}},\ }\href {\doibase
  10.1109/20.477575} {\bibfield  {journal} {\bibinfo  {journal} {IEEE Trans.
  Magn.}\ }\textbf {\bibinfo {volume} {32}},\ \bibinfo {pages} {230} (\bibinfo
  {year} {1996})}\BibitemShut {NoStop}%
\bibitem [{\citenamefont {Dreyer}\ \emph {et~al.}(2007)\citenamefont {Dreyer},
  \citenamefont {Norpoth}, \citenamefont {Jooss}, \citenamefont {Sievers},
  \citenamefont {Siegner}, \citenamefont {Neu},\ and\ \citenamefont
  {Johansen}}]{Dreyer2007}%
  \BibitemOpen
  \bibfield  {author} {\bibinfo {author} {\bibfnamefont {S.}~\bibnamefont
  {Dreyer}}, \bibinfo {author} {\bibfnamefont {J.}~\bibnamefont {Norpoth}},
  \bibinfo {author} {\bibfnamefont {C.}~\bibnamefont {Jooss}}, \bibinfo
  {author} {\bibfnamefont {S.}~\bibnamefont {Sievers}}, \bibinfo {author}
  {\bibfnamefont {U.}~\bibnamefont {Siegner}}, \bibinfo {author} {\bibfnamefont
  {V.}~\bibnamefont {Neu}}, \ and\ \bibinfo {author} {\bibfnamefont {T.~H.}\
  \bibnamefont {Johansen}},\ }\href {\doibase 10.1063/1.2717560} {\bibfield
  {journal} {\bibinfo  {journal} {J. Appl. Phys.}\ }\textbf {\bibinfo {volume}
  {101}},\ \bibinfo {pages} {083905} (\bibinfo {year} {2007})}\BibitemShut
  {NoStop}%
\bibitem [{\citenamefont {Anahory}\ \emph {et~al.}(2014)\citenamefont
  {Anahory}, \citenamefont {Reiner}, \citenamefont {Embon}, \citenamefont
  {Halbertal}, \citenamefont {Yakovenko}, \citenamefont {Myasoedov},
  \citenamefont {Rappaport}, \citenamefont {Huber},\ and\ \citenamefont
  {Zeldov}}]{Anahory2014}%
  \BibitemOpen
  \bibfield  {author} {\bibinfo {author} {\bibfnamefont {Y.}~\bibnamefont
  {Anahory}}, \bibinfo {author} {\bibfnamefont {J.}~\bibnamefont {Reiner}},
  \bibinfo {author} {\bibfnamefont {L.}~\bibnamefont {Embon}}, \bibinfo
  {author} {\bibfnamefont {D.}~\bibnamefont {Halbertal}}, \bibinfo {author}
  {\bibfnamefont {A.}~\bibnamefont {Yakovenko}}, \bibinfo {author}
  {\bibfnamefont {Y.}~\bibnamefont {Myasoedov}}, \bibinfo {author}
  {\bibfnamefont {M.~L.}\ \bibnamefont {Rappaport}}, \bibinfo {author}
  {\bibfnamefont {M.~E.}\ \bibnamefont {Huber}}, \ and\ \bibinfo {author}
  {\bibfnamefont {E.}~\bibnamefont {Zeldov}},\ }\href {\doibase
  10.1021/nl503022q} {\bibfield  {journal} {\bibinfo  {journal} {Nano Lett.}\
  }\textbf {\bibinfo {volume} {14}},\ \bibinfo {pages} {6481} (\bibinfo {year}
  {2014})}\BibitemShut {NoStop}%
\bibitem [{\citenamefont {Maertz}\ \emph {et~al.}(2010)\citenamefont {Maertz},
  \citenamefont {Wijnheijmer}, \citenamefont {Fuchs}, \citenamefont
  {Nowakowski},\ and\ \citenamefont {Awschalom}}]{Maertz2010}%
  \BibitemOpen
  \bibfield  {author} {\bibinfo {author} {\bibfnamefont {B.~J.}\ \bibnamefont
  {Maertz}}, \bibinfo {author} {\bibfnamefont {A.~P.}\ \bibnamefont
  {Wijnheijmer}}, \bibinfo {author} {\bibfnamefont {G.~D.}\ \bibnamefont
  {Fuchs}}, \bibinfo {author} {\bibfnamefont {M.~E.}\ \bibnamefont
  {Nowakowski}}, \ and\ \bibinfo {author} {\bibfnamefont {D.~D.}\ \bibnamefont
  {Awschalom}},\ }\href {\doibase 10.1063/1.3337096} {\bibfield  {journal}
  {\bibinfo  {journal} {Appl. Phys. Lett.}\ }\textbf {\bibinfo {volume} {96}},\
  \bibinfo {pages} {092504} (\bibinfo {year} {2010})}\BibitemShut {NoStop}%
\bibitem [{\citenamefont {Feldmann}(2004)}]{Feldmann2004}%
  \BibitemOpen
  \bibfield  {author} {\bibinfo {author} {\bibfnamefont {D.~M.}\ \bibnamefont
  {Feldmann}},\ }\href {\doibase 10.1103/PhysRevB.69.144515} {\bibfield
  {journal} {\bibinfo  {journal} {Phys. Rev. B}\ }\textbf {\bibinfo {volume}
  {69}},\ \bibinfo {pages} {144515} (\bibinfo {year} {2004})}\BibitemShut
  {NoStop}%
\bibitem [{\citenamefont {Tetienne}\ \emph {et~al.}(2019)\citenamefont
  {Tetienne}, \citenamefont {Dontschuk}, \citenamefont {Broadway},
  \citenamefont {Lillie}, \citenamefont {Teraji}, \citenamefont {Simpson},
  \citenamefont {Stacey},\ and\ \citenamefont {Hollenberg}}]{Tetienne2019}%
  \BibitemOpen
  \bibfield  {author} {\bibinfo {author} {\bibfnamefont {J.-P.}\ \bibnamefont
  {Tetienne}}, \bibinfo {author} {\bibfnamefont {N.}~\bibnamefont {Dontschuk}},
  \bibinfo {author} {\bibfnamefont {D.~A.}\ \bibnamefont {Broadway}}, \bibinfo
  {author} {\bibfnamefont {S.~E.}\ \bibnamefont {Lillie}}, \bibinfo {author}
  {\bibfnamefont {T.}~\bibnamefont {Teraji}}, \bibinfo {author} {\bibfnamefont
  {D.~A.}\ \bibnamefont {Simpson}}, \bibinfo {author} {\bibfnamefont
  {A.}~\bibnamefont {Stacey}}, \ and\ \bibinfo {author} {\bibfnamefont
  {L.~C.~L.}\ \bibnamefont {Hollenberg}},\ }\href {\doibase
  10.1103/PhysRevB.99.014436} {\bibfield  {journal} {\bibinfo  {journal} {Phys.
  Rev. B}\ }\textbf {\bibinfo {volume} {99}},\ \bibinfo {pages} {014436}
  (\bibinfo {year} {2019})}\BibitemShut {NoStop}%
\bibitem [{\citenamefont {Pelliccione}\ \emph {et~al.}(2016)\citenamefont
  {Pelliccione}, \citenamefont {Jenkins}, \citenamefont {Ovartchaiyapong},
  \citenamefont {Reetz}, \citenamefont {Emmanouilidou}, \citenamefont {Ni},\
  and\ \citenamefont {{Bleszynski Jayich}}}]{Pelliccione2016}%
  \BibitemOpen
  \bibfield  {author} {\bibinfo {author} {\bibfnamefont {M.}~\bibnamefont
  {Pelliccione}}, \bibinfo {author} {\bibfnamefont {A.}~\bibnamefont
  {Jenkins}}, \bibinfo {author} {\bibfnamefont {P.}~\bibnamefont
  {Ovartchaiyapong}}, \bibinfo {author} {\bibfnamefont {C.}~\bibnamefont
  {Reetz}}, \bibinfo {author} {\bibfnamefont {E.}~\bibnamefont
  {Emmanouilidou}}, \bibinfo {author} {\bibfnamefont {N.}~\bibnamefont {Ni}}, \
  and\ \bibinfo {author} {\bibfnamefont {A.~C.}\ \bibnamefont {{Bleszynski
  Jayich}}},\ }\href {\doibase 10.1038/nnano.2016.68} {\bibfield  {journal}
  {\bibinfo  {journal} {Nat. Nanotechnol.}\ }\textbf {\bibinfo {volume} {11}},\
  \bibinfo {pages} {700} (\bibinfo {year} {2016})}\BibitemShut {NoStop}%
\bibitem [{\citenamefont {Thiel}\ \emph {et~al.}(2016)\citenamefont {Thiel},
  \citenamefont {Rohner}, \citenamefont {Ganzhorn}, \citenamefont {Appel},
  \citenamefont {Neu}, \citenamefont {M{\"{u}}ller}, \citenamefont {Kleiner},
  \citenamefont {Koelle},\ and\ \citenamefont {Maletinsky}}]{Thiel2016}%
  \BibitemOpen
  \bibfield  {author} {\bibinfo {author} {\bibfnamefont {L.}~\bibnamefont
  {Thiel}}, \bibinfo {author} {\bibfnamefont {D.}~\bibnamefont {Rohner}},
  \bibinfo {author} {\bibfnamefont {M.}~\bibnamefont {Ganzhorn}}, \bibinfo
  {author} {\bibfnamefont {P.}~\bibnamefont {Appel}}, \bibinfo {author}
  {\bibfnamefont {E.}~\bibnamefont {Neu}}, \bibinfo {author} {\bibfnamefont
  {B.}~\bibnamefont {M{\"{u}}ller}}, \bibinfo {author} {\bibfnamefont
  {R.}~\bibnamefont {Kleiner}}, \bibinfo {author} {\bibfnamefont
  {D.}~\bibnamefont {Koelle}}, \ and\ \bibinfo {author} {\bibfnamefont
  {P.}~\bibnamefont {Maletinsky}},\ }\href {\doibase 10.1038/nnano.2016.63}
  {\bibfield  {journal} {\bibinfo  {journal} {Nat. Nanotechnol.}\ }\textbf
  {\bibinfo {volume} {11}},\ \bibinfo {pages} {677} (\bibinfo {year}
  {2016})}\BibitemShut {NoStop}%
\bibitem [{\citenamefont {Chang}\ \emph {et~al.}(2017)\citenamefont {Chang},
  \citenamefont {Eichler}, \citenamefont {Rhensius}, \citenamefont
  {Lorenzelli},\ and\ \citenamefont {Degen}}]{Chang2017}%
  \BibitemOpen
  \bibfield  {author} {\bibinfo {author} {\bibfnamefont {K.}~\bibnamefont
  {Chang}}, \bibinfo {author} {\bibfnamefont {A.}~\bibnamefont {Eichler}},
  \bibinfo {author} {\bibfnamefont {J.}~\bibnamefont {Rhensius}}, \bibinfo
  {author} {\bibfnamefont {L.}~\bibnamefont {Lorenzelli}}, \ and\ \bibinfo
  {author} {\bibfnamefont {C.~L.}\ \bibnamefont {Degen}},\ }\href {\doibase
  10.1021/acs.nanolett.6b05304} {\bibfield  {journal} {\bibinfo  {journal}
  {Nano Lett.}\ }\textbf {\bibinfo {volume} {17}},\ \bibinfo {pages} {2367}
  (\bibinfo {year} {2017})}\BibitemShut {NoStop}%
\bibitem [{\citenamefont {Ku}\ \emph {et~al.}()\citenamefont {Ku},
  \citenamefont {Zhou}, \citenamefont {Li}, \citenamefont {Shin}, \citenamefont
  {Shi}, \citenamefont {Burch}, \citenamefont {Zhang}, \citenamefont {Casola},
  \citenamefont {Taniguchi}, \citenamefont {Watanabe}, \citenamefont {Kim},
  \citenamefont {Yacoby},\ and\ \citenamefont {Walsworth}}]{Ku2019}%
  \BibitemOpen
  \bibfield  {author} {\bibinfo {author} {\bibfnamefont {M.~J.~H.}\
  \bibnamefont {Ku}}, \bibinfo {author} {\bibfnamefont {T.~X.}\ \bibnamefont
  {Zhou}}, \bibinfo {author} {\bibfnamefont {Q.}~\bibnamefont {Li}}, \bibinfo
  {author} {\bibfnamefont {Y.~J.}\ \bibnamefont {Shin}}, \bibinfo {author}
  {\bibfnamefont {J.~K.}\ \bibnamefont {Shi}}, \bibinfo {author} {\bibfnamefont
  {C.}~\bibnamefont {Burch}}, \bibinfo {author} {\bibfnamefont
  {H.}~\bibnamefont {Zhang}}, \bibinfo {author} {\bibfnamefont
  {F.}~\bibnamefont {Casola}}, \bibinfo {author} {\bibfnamefont
  {T.}~\bibnamefont {Taniguchi}}, \bibinfo {author} {\bibfnamefont
  {K.}~\bibnamefont {Watanabe}}, \bibinfo {author} {\bibfnamefont
  {P.}~\bibnamefont {Kim}}, \bibinfo {author} {\bibfnamefont {A.}~\bibnamefont
  {Yacoby}}, \ and\ \bibinfo {author} {\bibfnamefont {R.~L.}\ \bibnamefont
  {Walsworth}},\ }\href {http://arxiv.org/abs/1905.10791} {\ }\Eprint
  {http://arxiv.org/abs/1905.10791} {arXiv:1905.10791} \BibitemShut {NoStop}%
\bibitem [{\citenamefont {Lillie}\ \emph {et~al.}(2020)\citenamefont {Lillie},
  \citenamefont {Broadway}, \citenamefont {Dontschuk}, \citenamefont
  {Scholten}, \citenamefont {Johnson}, \citenamefont {Wolf}, \citenamefont
  {Rachel}, \citenamefont {Hollenberg},\ and\ \citenamefont
  {Tetienne}}]{Lillie2020}%
  \BibitemOpen
  \bibfield  {author} {\bibinfo {author} {\bibfnamefont {S.~E.}\ \bibnamefont
  {Lillie}}, \bibinfo {author} {\bibfnamefont {D.~A.}\ \bibnamefont
  {Broadway}}, \bibinfo {author} {\bibfnamefont {N.}~\bibnamefont {Dontschuk}},
  \bibinfo {author} {\bibfnamefont {S.~C.}\ \bibnamefont {Scholten}}, \bibinfo
  {author} {\bibfnamefont {B.~C.}\ \bibnamefont {Johnson}}, \bibinfo {author}
  {\bibfnamefont {S.}~\bibnamefont {Wolf}}, \bibinfo {author} {\bibfnamefont
  {S.}~\bibnamefont {Rachel}}, \bibinfo {author} {\bibfnamefont {L.~C.~L.}\
  \bibnamefont {Hollenberg}}, \ and\ \bibinfo {author} {\bibfnamefont {J.-P.}\
  \bibnamefont {Tetienne}},\ }\href {\doibase 10.1021/acs.nanolett.9b05071}
  {\bibfield  {journal} {\bibinfo  {journal} {Nano Lett.}\ }\textbf {\bibinfo
  {volume} {20}},\ \bibinfo {pages} {1855} (\bibinfo {year}
  {2020})}\BibitemShut {NoStop}%
\bibitem [{\citenamefont {Tetienne}\ \emph {et~al.}(2018)\citenamefont
  {Tetienne}, \citenamefont {{De Gille}}, \citenamefont {Broadway},
  \citenamefont {Teraji}, \citenamefont {Lillie}, \citenamefont {McCoey},
  \citenamefont {Dontschuk}, \citenamefont {Hall}, \citenamefont {Stacey},
  \citenamefont {Simpson},\ and\ \citenamefont {Hollenberg}}]{Tetienne2018a}%
  \BibitemOpen
  \bibfield  {author} {\bibinfo {author} {\bibfnamefont {J.-P.}\ \bibnamefont
  {Tetienne}}, \bibinfo {author} {\bibfnamefont {R.~W.}\ \bibnamefont {{De
  Gille}}}, \bibinfo {author} {\bibfnamefont {D.~A.}\ \bibnamefont {Broadway}},
  \bibinfo {author} {\bibfnamefont {T.}~\bibnamefont {Teraji}}, \bibinfo
  {author} {\bibfnamefont {S.~E.}\ \bibnamefont {Lillie}}, \bibinfo {author}
  {\bibfnamefont {J.~M.}\ \bibnamefont {McCoey}}, \bibinfo {author}
  {\bibfnamefont {N.}~\bibnamefont {Dontschuk}}, \bibinfo {author}
  {\bibfnamefont {L.~T.}\ \bibnamefont {Hall}}, \bibinfo {author}
  {\bibfnamefont {A.}~\bibnamefont {Stacey}}, \bibinfo {author} {\bibfnamefont
  {D.~A.}\ \bibnamefont {Simpson}}, \ and\ \bibinfo {author} {\bibfnamefont
  {L.~C.~L.}\ \bibnamefont {Hollenberg}},\ }\href {\doibase
  10.1103/PhysRevB.97.085402} {\bibfield  {journal} {\bibinfo  {journal} {Phys.
  Rev. B}\ }\textbf {\bibinfo {volume} {97}},\ \bibinfo {pages} {85402}
  (\bibinfo {year} {2018})}\BibitemShut {NoStop}%
\bibitem [{\citenamefont {Broadway}\ \emph {et~al.}(2018)\citenamefont
  {Broadway}, \citenamefont {Dontschuk}, \citenamefont {Tsai}, \citenamefont
  {Lillie}, \citenamefont {Lew}, \citenamefont {McCallum}, \citenamefont
  {Johnson}, \citenamefont {Doherty}, \citenamefont {Stacey}, \citenamefont
  {Hollenberg},\ and\ \citenamefont {Tetienne}}]{Broadway2018b}%
  \BibitemOpen
  \bibfield  {author} {\bibinfo {author} {\bibfnamefont {D.~A.}\ \bibnamefont
  {Broadway}}, \bibinfo {author} {\bibfnamefont {N.}~\bibnamefont {Dontschuk}},
  \bibinfo {author} {\bibfnamefont {A.}~\bibnamefont {Tsai}}, \bibinfo {author}
  {\bibfnamefont {S.~E.}\ \bibnamefont {Lillie}}, \bibinfo {author}
  {\bibfnamefont {C.~T.-K.}\ \bibnamefont {Lew}}, \bibinfo {author}
  {\bibfnamefont {J.~C.}\ \bibnamefont {McCallum}}, \bibinfo {author}
  {\bibfnamefont {B.~C.}\ \bibnamefont {Johnson}}, \bibinfo {author}
  {\bibfnamefont {M.~W.}\ \bibnamefont {Doherty}}, \bibinfo {author}
  {\bibfnamefont {A.}~\bibnamefont {Stacey}}, \bibinfo {author} {\bibfnamefont
  {L.~C.~L.}\ \bibnamefont {Hollenberg}}, \ and\ \bibinfo {author}
  {\bibfnamefont {J.-P.}\ \bibnamefont {Tetienne}},\ }\href {\doibase
  10.1038/s41928-018-0130-0} {\bibfield  {journal} {\bibinfo  {journal} {Nat.
  Electron.}\ }\textbf {\bibinfo {volume} {1}},\ \bibinfo {pages} {502}
  (\bibinfo {year} {2018})}\BibitemShut {NoStop}%
\bibitem [{\citenamefont {Broadway}\ \emph {et~al.}(2019)\citenamefont
  {Broadway}, \citenamefont {Johnson}, \citenamefont {Barson}, \citenamefont
  {Lillie}, \citenamefont {Dontschuk}, \citenamefont {McCloskey}, \citenamefont
  {Tsai}, \citenamefont {Teraji}, \citenamefont {Simpson}, \citenamefont
  {Stacey}, \citenamefont {McCallum}, \citenamefont {Bradby}, \citenamefont
  {Doherty}, \citenamefont {Hollenberg},\ and\ \citenamefont
  {Tetienne}}]{Broadway2018c}%
  \BibitemOpen
  \bibfield  {author} {\bibinfo {author} {\bibfnamefont {D.~A.}\ \bibnamefont
  {Broadway}}, \bibinfo {author} {\bibfnamefont {B.~C.}\ \bibnamefont
  {Johnson}}, \bibinfo {author} {\bibfnamefont {M.~S.~J.}\ \bibnamefont
  {Barson}}, \bibinfo {author} {\bibfnamefont {S.~E.}\ \bibnamefont {Lillie}},
  \bibinfo {author} {\bibfnamefont {N.}~\bibnamefont {Dontschuk}}, \bibinfo
  {author} {\bibfnamefont {D.~J.}\ \bibnamefont {McCloskey}}, \bibinfo {author}
  {\bibfnamefont {A.}~\bibnamefont {Tsai}}, \bibinfo {author} {\bibfnamefont
  {T.}~\bibnamefont {Teraji}}, \bibinfo {author} {\bibfnamefont {D.~A.}\
  \bibnamefont {Simpson}}, \bibinfo {author} {\bibfnamefont {A.}~\bibnamefont
  {Stacey}}, \bibinfo {author} {\bibfnamefont {J.~C.}\ \bibnamefont
  {McCallum}}, \bibinfo {author} {\bibfnamefont {J.~E.}\ \bibnamefont
  {Bradby}}, \bibinfo {author} {\bibfnamefont {M.~W.}\ \bibnamefont {Doherty}},
  \bibinfo {author} {\bibfnamefont {L.~C.~L.}\ \bibnamefont {Hollenberg}}, \
  and\ \bibinfo {author} {\bibfnamefont {J.-P.}\ \bibnamefont {Tetienne}},\
  }\href {\doibase 10.1021/acs.nanolett.9b01402} {\bibfield  {journal}
  {\bibinfo  {journal} {Nano Lett.}\ }\textbf {\bibinfo {volume} {19}},\
  \bibinfo {pages} {4543} (\bibinfo {year} {2019})}\BibitemShut {NoStop}%
\bibitem [{\citenamefont {Simpson}\ \emph {et~al.}(2016)\citenamefont
  {Simpson}, \citenamefont {Tetienne}, \citenamefont {McCoey}, \citenamefont
  {Ganesan}, \citenamefont {Hall}, \citenamefont {Petrou}, \citenamefont
  {Scholten},\ and\ \citenamefont {Hollenberg}}]{Simpson2016}%
  \BibitemOpen
  \bibfield  {author} {\bibinfo {author} {\bibfnamefont {D.~A.}\ \bibnamefont
  {Simpson}}, \bibinfo {author} {\bibfnamefont {J.~P.}\ \bibnamefont
  {Tetienne}}, \bibinfo {author} {\bibfnamefont {J.~M.}\ \bibnamefont
  {McCoey}}, \bibinfo {author} {\bibfnamefont {K.}~\bibnamefont {Ganesan}},
  \bibinfo {author} {\bibfnamefont {L.~T.}\ \bibnamefont {Hall}}, \bibinfo
  {author} {\bibfnamefont {S.}~\bibnamefont {Petrou}}, \bibinfo {author}
  {\bibfnamefont {R.~E.}\ \bibnamefont {Scholten}}, \ and\ \bibinfo {author}
  {\bibfnamefont {L.~C.}\ \bibnamefont {Hollenberg}},\ }\href {\doibase
  10.1038/srep22797} {\bibfield  {journal} {\bibinfo  {journal} {Sci. Rep.}\
  }\textbf {\bibinfo {volume} {6}},\ \bibinfo {pages} {22797} (\bibinfo {year}
  {2016})}\BibitemShut {NoStop}%
\bibitem [{\citenamefont {{Le Sage}}\ \emph {et~al.}(2013)\citenamefont {{Le
  Sage}}, \citenamefont {Arai}, \citenamefont {Glenn}, \citenamefont
  {DeVience}, \citenamefont {Pham}, \citenamefont {Rahn-Lee}, \citenamefont
  {Lukin}, \citenamefont {Yacoby}, \citenamefont {Komeili},\ and\ \citenamefont
  {Walsworth}}]{LeSage2013}%
  \BibitemOpen
  \bibfield  {author} {\bibinfo {author} {\bibfnamefont {D.}~\bibnamefont {{Le
  Sage}}}, \bibinfo {author} {\bibfnamefont {K.}~\bibnamefont {Arai}}, \bibinfo
  {author} {\bibfnamefont {D.~R.}\ \bibnamefont {Glenn}}, \bibinfo {author}
  {\bibfnamefont {S.~J.}\ \bibnamefont {DeVience}}, \bibinfo {author}
  {\bibfnamefont {L.~M.}\ \bibnamefont {Pham}}, \bibinfo {author}
  {\bibfnamefont {L.}~\bibnamefont {Rahn-Lee}}, \bibinfo {author}
  {\bibfnamefont {M.~D.}\ \bibnamefont {Lukin}}, \bibinfo {author}
  {\bibfnamefont {A.}~\bibnamefont {Yacoby}}, \bibinfo {author} {\bibfnamefont
  {A.}~\bibnamefont {Komeili}}, \ and\ \bibinfo {author} {\bibfnamefont
  {R.~L.}\ \bibnamefont {Walsworth}},\ }\href {\doibase 10.1038/nature12072}
  {\bibfield  {journal} {\bibinfo  {journal} {Nature}\ }\textbf {\bibinfo
  {volume} {496}},\ \bibinfo {pages} {486} (\bibinfo {year}
  {2013})}\BibitemShut {NoStop}%
\bibitem [{\citenamefont {Levine}\ \emph {et~al.}()\citenamefont {Levine},
  \citenamefont {Turner}, \citenamefont {Kehayias}, \citenamefont {Hart},
  \citenamefont {Langellier}, \citenamefont {Trubko}, \citenamefont {Glenn},
  \citenamefont {Fu},\ and\ \citenamefont {Walsworth}}]{Levine}%
  \BibitemOpen
  \bibfield  {author} {\bibinfo {author} {\bibfnamefont {E.~V.}\ \bibnamefont
  {Levine}}, \bibinfo {author} {\bibfnamefont {M.~J.}\ \bibnamefont {Turner}},
  \bibinfo {author} {\bibfnamefont {P.}~\bibnamefont {Kehayias}}, \bibinfo
  {author} {\bibfnamefont {C.~A.}\ \bibnamefont {Hart}}, \bibinfo {author}
  {\bibfnamefont {N.}~\bibnamefont {Langellier}}, \bibinfo {author}
  {\bibfnamefont {R.}~\bibnamefont {Trubko}}, \bibinfo {author} {\bibfnamefont
  {D.~R.}\ \bibnamefont {Glenn}}, \bibinfo {author} {\bibfnamefont {R.~R.}\
  \bibnamefont {Fu}}, \ and\ \bibinfo {author} {\bibfnamefont {R.~L.}\
  \bibnamefont {Walsworth}},\ }\href {http://arxiv.org/abs/1910.00061} {\
  }\Eprint {http://arxiv.org/abs/1910.00061} {arXiv:1910.00061} \BibitemShut
  {NoStop}%
\bibitem [{\citenamefont {Rohner}\ \emph {et~al.}(2019)\citenamefont {Rohner},
  \citenamefont {Happacher}, \citenamefont {Reiser}, \citenamefont {Tschudin},
  \citenamefont {Tallaire}, \citenamefont {Achard}, \citenamefont {Shields},\
  and\ \citenamefont {Maletinsky}}]{Rohner2019}%
  \BibitemOpen
  \bibfield  {author} {\bibinfo {author} {\bibfnamefont {D.}~\bibnamefont
  {Rohner}}, \bibinfo {author} {\bibfnamefont {J.}~\bibnamefont {Happacher}},
  \bibinfo {author} {\bibfnamefont {P.}~\bibnamefont {Reiser}}, \bibinfo
  {author} {\bibfnamefont {M.~A.}\ \bibnamefont {Tschudin}}, \bibinfo {author}
  {\bibfnamefont {A.}~\bibnamefont {Tallaire}}, \bibinfo {author}
  {\bibfnamefont {J.}~\bibnamefont {Achard}}, \bibinfo {author} {\bibfnamefont
  {B.~J.}\ \bibnamefont {Shields}}, \ and\ \bibinfo {author} {\bibfnamefont
  {P.}~\bibnamefont {Maletinsky}},\ }\href {\doibase 10.1063/1.5127101}
  {\bibfield  {journal} {\bibinfo  {journal} {Appl. Phys. Lett.}\ }\textbf
  {\bibinfo {volume} {115}},\ \bibinfo {pages} {192401} (\bibinfo {year}
  {2019})}\BibitemShut {NoStop}%
\bibitem [{\citenamefont {Tian}\ \emph {et~al.}(2019)\citenamefont {Tian},
  \citenamefont {Zhang}, \citenamefont {Li}, \citenamefont {Ying},
  \citenamefont {Li}, \citenamefont {Zhang}, \citenamefont {Liu},\ and\
  \citenamefont {Lei}}]{Tian2019}%
  \BibitemOpen
  \bibfield  {author} {\bibinfo {author} {\bibfnamefont {S.}~\bibnamefont
  {Tian}}, \bibinfo {author} {\bibfnamefont {J.-F.}\ \bibnamefont {Zhang}},
  \bibinfo {author} {\bibfnamefont {C.}~\bibnamefont {Li}}, \bibinfo {author}
  {\bibfnamefont {T.}~\bibnamefont {Ying}}, \bibinfo {author} {\bibfnamefont
  {S.}~\bibnamefont {Li}}, \bibinfo {author} {\bibfnamefont {X.}~\bibnamefont
  {Zhang}}, \bibinfo {author} {\bibfnamefont {K.}~\bibnamefont {Liu}}, \ and\
  \bibinfo {author} {\bibfnamefont {H.}~\bibnamefont {Lei}},\ }\href {\doibase
  10.1021/jacs.8b13584} {\bibfield  {journal} {\bibinfo  {journal} {J. Am.
  Chem. Soc.}\ }\textbf {\bibinfo {volume} {141}},\ \bibinfo {pages} {5326}
  (\bibinfo {year} {2019})}\BibitemShut {NoStop}%
\bibitem [{\citenamefont {Son}\ \emph {et~al.}(2019)\citenamefont {Son},
  \citenamefont {Coak}, \citenamefont {Lee}, \citenamefont {Kim}, \citenamefont
  {Kim}, \citenamefont {Hamidov}, \citenamefont {Cho}, \citenamefont {Liu},
  \citenamefont {Jarvis}, \citenamefont {Brown}, \citenamefont {Kim},
  \citenamefont {Park}, \citenamefont {Khomskii}, \citenamefont {Saxena},\ and\
  \citenamefont {Park}}]{Son2019}%
  \BibitemOpen
  \bibfield  {author} {\bibinfo {author} {\bibfnamefont {S.}~\bibnamefont
  {Son}}, \bibinfo {author} {\bibfnamefont {M.~J.}\ \bibnamefont {Coak}},
  \bibinfo {author} {\bibfnamefont {N.}~\bibnamefont {Lee}}, \bibinfo {author}
  {\bibfnamefont {J.}~\bibnamefont {Kim}}, \bibinfo {author} {\bibfnamefont
  {T.~Y.}\ \bibnamefont {Kim}}, \bibinfo {author} {\bibfnamefont
  {H.}~\bibnamefont {Hamidov}}, \bibinfo {author} {\bibfnamefont
  {H.}~\bibnamefont {Cho}}, \bibinfo {author} {\bibfnamefont {C.}~\bibnamefont
  {Liu}}, \bibinfo {author} {\bibfnamefont {D.~M.}\ \bibnamefont {Jarvis}},
  \bibinfo {author} {\bibfnamefont {P.~A.}\ \bibnamefont {Brown}}, \bibinfo
  {author} {\bibfnamefont {J.~H.}\ \bibnamefont {Kim}}, \bibinfo {author}
  {\bibfnamefont {C.~H.}\ \bibnamefont {Park}}, \bibinfo {author}
  {\bibfnamefont {D.~I.}\ \bibnamefont {Khomskii}}, \bibinfo {author}
  {\bibfnamefont {S.~S.}\ \bibnamefont {Saxena}}, \ and\ \bibinfo {author}
  {\bibfnamefont {J.~G.}\ \bibnamefont {Park}},\ }\href {\doibase
  10.1103/PhysRevB.99.041402} {\bibfield  {journal} {\bibinfo  {journal} {Phys.
  Rev. B}\ }\textbf {\bibinfo {volume} {99}},\ \bibinfo {pages} {1} (\bibinfo
  {year} {2019})}\BibitemShut {NoStop}%
\bibitem [{\citenamefont {Kong}\ \emph {et~al.}(2019)\citenamefont {Kong},
  \citenamefont {Stolze}, \citenamefont {Timmons}, \citenamefont {Tao},
  \citenamefont {Ni}, \citenamefont {Guo}, \citenamefont {Yang}, \citenamefont
  {Prozorov},\ and\ \citenamefont {Cava}}]{Kong2019}%
  \BibitemOpen
  \bibfield  {author} {\bibinfo {author} {\bibfnamefont {T.}~\bibnamefont
  {Kong}}, \bibinfo {author} {\bibfnamefont {K.}~\bibnamefont {Stolze}},
  \bibinfo {author} {\bibfnamefont {E.~I.}\ \bibnamefont {Timmons}}, \bibinfo
  {author} {\bibfnamefont {J.}~\bibnamefont {Tao}}, \bibinfo {author}
  {\bibfnamefont {D.}~\bibnamefont {Ni}}, \bibinfo {author} {\bibfnamefont
  {S.}~\bibnamefont {Guo}}, \bibinfo {author} {\bibfnamefont {Z.}~\bibnamefont
  {Yang}}, \bibinfo {author} {\bibfnamefont {R.}~\bibnamefont {Prozorov}}, \
  and\ \bibinfo {author} {\bibfnamefont {R.~J.}\ \bibnamefont {Cava}},\ }\href
  {\doibase 10.1002/adma.201808074} {\bibfield  {journal} {\bibinfo  {journal}
  {Adv. Mater.}\ }\textbf {\bibinfo {volume} {31}},\ \bibinfo {pages} {1}
  (\bibinfo {year} {2019})}\BibitemShut {NoStop}%
\end{thebibliography}%

\end{document}